\documentclass[onecolumn]{mn2e}
\usepackage{graphicx}

\def\pmb#1{\setbox0=\hbox{#1}%
    \kern-.025em\copy0\kern-\wd0
    \kern.05em\copy0\kern-\wd0
    \kern-.025em\raise.0433em\box0 }
\def\btheta{\pmb{$\theta$}}

\newcounter{parentequation}
\def\beglet{
  \addtocounter{equation}{1}%
  \setcounter{parentequation}{\value{equation}}%
  \setcounter{equation}{0}%
  \def\theequation{\arabic{parentequation}\alph{equation}}%
  \ignorespaces
}
\def\endlet{
  \setcounter{equation}{\value{parentequation}}%
  \def\theequation{\arabic{equation}}%
}

\def\ltsima{$\; \buildrel < \over \sim \;$}
\def\gtsima{$\; \buildrel > \over \sim \;$}
\def\simlt{\lower.5ex\hbox{\ltsima}}
\def\simgt{\lower.5ex\hbox{\gtsima}}
\def\kms{{\rm kms}^{-1}}

\def\etal{{\it et al.}\rm}
\def\etals{{\it et al. }\rm}
\def\kmsMpc{{\rm km}\ {\rm s}^{-1}{\rm Mpc}^{-1}}
\def\planck{{\it Planck}}
\def\Planck{{\it Planck}}
\def\LCDM{$\Lambda$CDM}
\def\lcdm{$\Lambda$CDM}

\begin{document}

\title[$H_0$ Revisited]
{$H_0$ Revisited}

\author[G. Efstathiou]{George Efstathiou\\
 Kavli Institute for Cosmology and Institute of Astronomy, Madingley Road, Cambridge, CB3 OHA.}

\maketitle

\begin{abstract}
  I reanalyse the Riess {\it et al.} (2011, hereafter R11) Cepheid
  data using the revised geometric maser distance to NGC 4258 of
  Humphreys {\it et al.} (2013, hereafter H13). I explore different
  outlier rejection criteria designed to give a reduced $\chi^2$ of
  unity and compare the results with the R11 rejection algorithm,
  which produces a reduced $\chi^2$ that is substantially less than
  unity and, in some cases, leads to underestimates of the errors on
  parameters.  I show that there are sub-luminous low metallicity
  Cepheids in the R11 sample that skew the global fits of the
  period-luminosity relation. This has a small but non-negligible
  impact on the global fits using NGC 4258 as a distance scale anchor,
  but adds a poorly constrained source of systematic error when using
  the Large Magellanic Cloud (LMC) as an anchor. I also show that the
  small Milky Way (MW) Cepheid sample with accurate parallax
  measurements leads to a distance to NGC 4258 that is in tension with
  the maser distance. I conclude that $H_0$ based on the NGC 4258
  maser distance is $H_0 = 70.6 \pm 3.3 \ \kmsMpc$, compatible within
  $1\sigma$ with the recent determination from \planck\ for the base
  six-parameter \LCDM\ cosmology. If the H-band period-luminosity
  relation is assumed to be independent of metallicity and the three
  distance anchors are combined, I find $H_0 = 72.5 \pm 2.5 \
  \kmsMpc$, which differs by $1.9\sigma$ from the \planck\ value. The
  differences between the \planck\ results and these estimates of
  $H_0$ are not large enough to provide compelling evidence for new
  physics at this stage.

\vskip 0.15 truein

\noindent
{\bf Key words}: cosmology: distance scale,  cosmological parameters.

\vskip 0.1 truein

\end{abstract}

\section{Introduction}

The recent \Planck\ observations of the cosmic microwave background
(CMB) lead to a Hubble constant of $H_0 = 67.3 \pm 1.2 \ \kmsMpc$ for
the base six-parameter $\Lambda$CDM model (Planck Collaboration 2013,
herafter P13).  This value is in tension, at about the $2.5 \sigma$
level, with the direct measurement of $H_0 = 73.8 \pm 2.4 \ \kmsMpc$
reported by R11.  If these numbers are taken at face value, they
suggest evidence for new physics at about the $2.5 \sigma$ level (for
example, exotic physics in the neutrino or dark energy sectors as
discussed in P13; see also Wyman \etals 2013, Hamann and Hasenkamp
2013; Battye and Moss 2013; Rest \etals 2013; Suyu \etals 2013). The exciting possibility
of discovering new physics provides strong motivation to subject both the CMB and
$H_0$ measurements to intense scrutiny.

Direct astrophysical measurements of the Hubble constant have  a
checkered history (see, for example, the reviews by Tammann, Sandage
and Reindl 2008; Freedman and Madore 2010). The Hubble Space Telescope
(HST) Key Project led to a significant improvement in the control of
systematic errors leading to `final' estimate of $H_0= 72 \pm 8
\ \kmsMpc$ (Freedman \etals 2001).  Since then, two Cepheid based
programmes have been underway with the aim of reducing the error on
$H_0$: the {\it Supernovae and $H_0$ for the Equation of State}
(SH0ES) programme of R11 (with earlier results reported in Riess
\etals 2009) and the {\it Carnegie Hubble Program} of Freedman \etals
(2012). In addition, other programmes are underway using
geometrical methods, for example the Megamaser Cosmology Project (MCP)
(Reid \etals 2013; Braatz \etals 2013) and the Cosmological Monitoring
of Gravitational Lenses (COSMOGRAIL) project (Suyu \etals 2010, Courbin \etals 2011;
Trewes \etals 2013).

This paper presents a  reanalysis of the R11 Cepheid data. The $H_0$ measurement
from these data has the smallest error and has  been used widely in
combination with CMB measurements for cosmological parameter analysis
({\it e.g.}  Hinshaw \etals 2012; Hou \etals 2012; Sievers \etals
2013). The study reported  here was motivated by certain aspects of the R11
analysis:  the R11 outlier rejection algorithm (which rejects a
large fraction, $\sim 20\%$, of the Cepheids), the low reduced $\chi^2$ values
of their fits,  and  the variations of some of the parameter values 
with  different distance anchors,
particularly the metallicity dependence of the period-luminosity relation.

The layout of this paper is as follows. Section 2 reviews the near-IR
period-luminosity (P-L) relation of a sample of LMC Cepheids. Section
3 describes a reanalysis of the R11 sample using the maser distance to
NGC 4258 as an anchor.    Section 4 investigates the use of the
LMC and MW Cepheids as anchors. Section 5 investigates combinations of
the three distance anchors and presents some internal consistency
checks. The conclusions are summarized in
Section 6.

\section{The LMC Cepheids}

I will start with the LMC Cepheids which I will use as a reference since the slope
of the P-L relation is tightly constrained from this sample.
I use the  53 LMC Cepheids with H-band magnitudes listed in Persson \etals (2004) and
V, I magnitudes listed in Sebo \etals (2002).  Figure 1 shows the Wesenheit magnitudes
\begin{equation}
m_{W} = m_{H} - 0.41(V-I), \label{LMC1}
\end{equation}
plotted against period $P$ (in units of days). The line shows a
least-squares fit to
\begin{equation}
m^P_{W} = A + b_W \ ({\rm log} P - 1), \label{LMC2}
\end{equation}
{\it i.e.} minimising 
\begin{equation}
\chi^2 = \sum_i {(m_{W, i} - m_{W}^P)^2 \over (\sigma^2_{e, i} + \sigma^2_{\rm int})}, \label{LMC3}
\end{equation}
where $\sigma_{e, i}$ is the error on $m_{W, i}$ and $\sigma_{\rm int}$ is the `internal'
scatter that gives a reduced $\chi^2$ (denoted $\hat \chi^2$ in this paper) of unity. Since
$\sigma_{\rm int}$ depends on the  parameters $A$ and $b_W$, the minimisation is
performed iteratively until convergence.
The best fit parameters are
\beglet
\begin{eqnarray}
A & =&  12.555 \pm 0.018, \quad b_W = -3.23 \pm 0.06, \quad \sigma_{\rm int}=0.113,  \quad {\rm log}P < 1.8, \\ \label{LMC3a}
A & =&  12.594 \pm 0.034, \quad b_W = -3.35 \pm 0.11, \quad \sigma_{\rm int}=0.104,  \quad 1.0 < {\rm log}P < 1.8.  \label{LMC3b}
\end{eqnarray}
\endlet
The upper period limit is imposed because there is evidence that
the P-L relation departs from a power law for periods $\simgt 60$ days
(Persson \etal, 2004; Freedman \etals 2011; Scowcroft \etal 2011). The
lower period limit in (\ref{LMC3b}) has been imposed because there
have been some claims that the P-L relation at optical and near-IR
wavelengths changes at periods less than 10 days (Ngeow \etals
2009). The results of (4a) and (4b) and Figure 1 show no evidence for
any significant change in the power-law slope at low periods, in
agreement with the results of Persson \etals (2004). Evidently, a
single power law is an extremely good fit to the LMC Cepheids at least
to periods of 60 days, and the slope of the P-L relation is determined
to high accuracy.

\begin{figure*}

\includegraphics[width=100mm, angle=0]{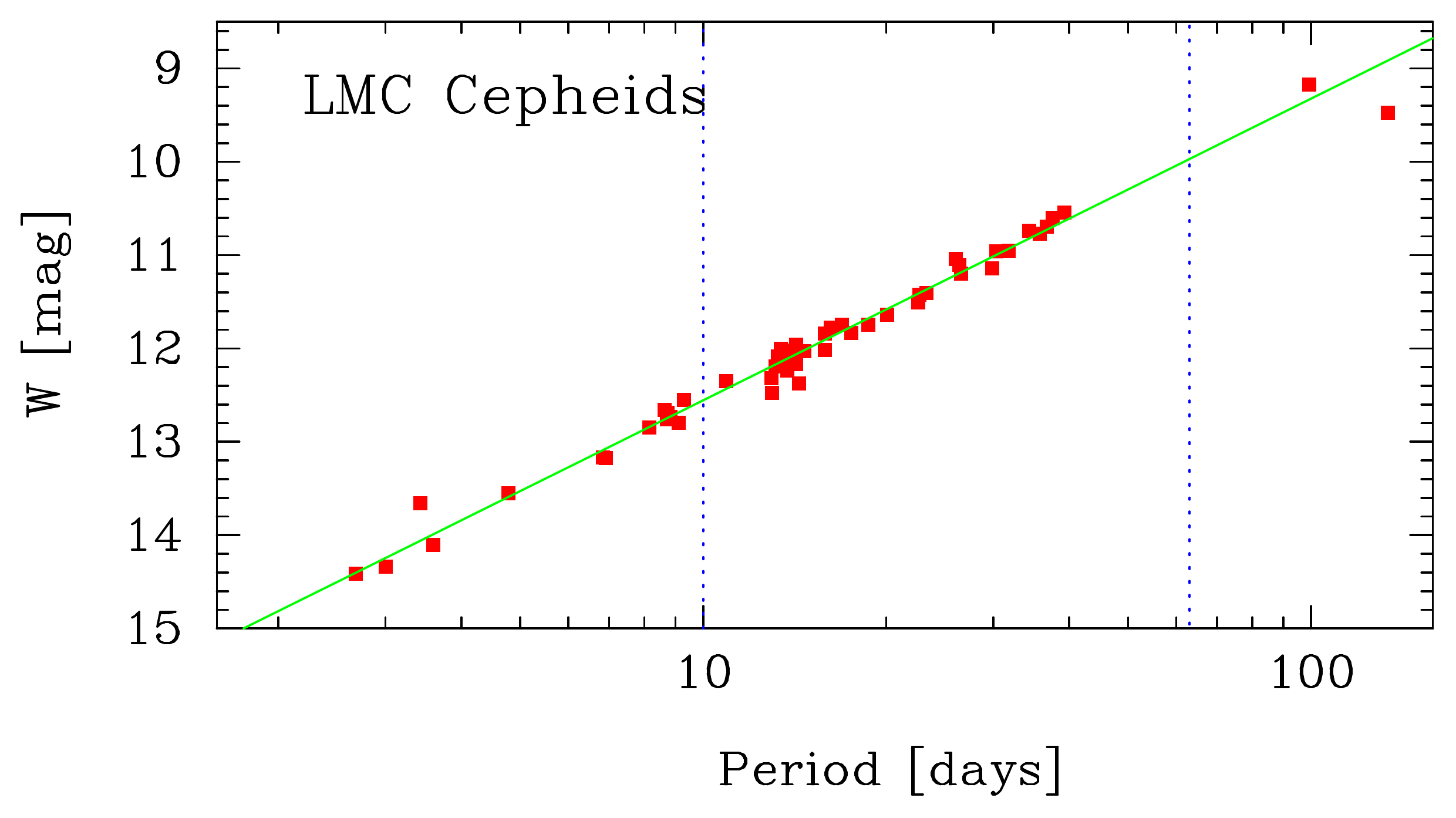}
\caption
{Period-luminosity relation for the LMC Cepheids. The line shows the best
fit of equation  (4a). The vertical  dotted lines show
the range of periods used in the fit of equation  (\ref{LMC3b}).}
\end{figure*}

\section{Analysis of the R11 Cepheid sample using NGC 4258 as an anchor distance}

\subsection{Outlier rejection}

As discussed by R11, there are several reasons to expect outliers in the P-L relation.
These include variables  misidentified as classical Cepheids, blended images, errors in
crowding corrections and possible aliasing  of the periods.

The R11 rejection algorithm works as follows:

\noindent
$\bullet$ The H-band only P-L relations are fitted galaxy-by-galaxy to a power law with  slope
fixed at $b_H=-3.1$ in the first iteration,  weighted by the magnitude errors in Table 2 of R11. 
Cepheids with periods $> 205$ days are excluded.

\noindent
$\bullet$  Cepheids are rejected
if they deviate from the best-fit relation by $\ge 0.75 \ {\rm mag}$, or by more than 
2.5 times the magnitude error.

\noindent
$\bullet$ The fitting and rejection is repeated iteratively 6 times.

Once the outliers have been removed, R11 proceed to global fits using
all of their galaxies, now adding a $0.21$ mag. error in quadrature to
the magnitude errors listed in their Table 2 (Riess, private communication).  
 One of the consequences of adding this additional error term is that the $\hat \chi^2$ values of
the R11 global fits are always less than unity (with typical values of $\hat \chi^2
\sim 0.65$) so R11 rescale their
covariance matrices by $1/\hat \chi^2$ to compute errors on
parameters.

\begin{table*}
\begin{center}

\begin{tabular}{lllllll} \hline
               &             &              & \multicolumn{2}{c}{H mags} &  \multicolumn{2}{c}{W mags}  \\
Galaxy         & $N_{\rm fit}$ & $N_{\rm rej}$ & $\sigma$ (mag) & $\hat \chi^2$ & $\sigma$ (mag) & $\hat\chi^2$\\ \hline
N4536         & 69 & 20 & $0.32$ & $1.25$ & $0.33$ & $1.22$   \\
N4639         & 32 &  6 & $0.41$ & $1.25$ & $0.41$ & $1.24$  \\
N3370         & 79 & 10 & $0.32$ & $0.90$ & $0.32$ & $0.89$  \\ 
N3982         & 29 & 22 & $0.32$ & $0.79$ & $0.33$ & $0.84$  \\
N3021         & 26 &  4 & $0.39$ & $0.73$ & $0.39$ & $0.74$  \\
N1309         & 36 &  4 & $0.31$ & $0.95$ & $0.32$ & $0.98$ \\
N5584         & 95 & 17 & $0.32$ & $1.01$ & $0.32$ & $1.04$ \\
N4038         & 39 &  7 & $0.34$ & $0.77$ & $0.36$ & $0.97$ \\
N4258         &164 & 46 & $0.36$ & $1.01$ & $0.37$ & $0.98$ \cr \hline
\end{tabular}
\caption{H-band rejection. $N_{\rm fit}$ and $N_{\rm rej}$ lists the number of 
Cepheids accepted and rejected by the R11 outlier rejection
algorithm. $\sigma$ is the standard deviation of the magnitude residuals about the best
fit P-L relation with slope constrained to $b=-3.1$ and $\hat \chi^2$ lists the reduced
$\chi^2$ for each fit (with no additional contribution from `intrinsic' scatter). 
Results are given for fits to H-band and Wesenheit P-L relations. }
\end{center}
\end{table*}

There are several aspects about this rejection algorithm that are worrisome:

\smallskip

\noindent
$\bullet$ The rejection algorithm is applied galaxy-by-galaxy {\it before} the global
fit to the entire sample.

\noindent

\noindent
$\bullet$ A large fraction of the data are rejected (about 20\% of the
total sample).

\noindent
$\bullet$ The imposition of an absolute cut of $0.75$ mag will accept
points with large magnitude errors and small residuals, {\it i.e.}
points that just happen to lie close to the best-fit P-L relation for
each galaxy.

\noindent
$\bullet$ As a consequence, $\hat \chi^2$ for the {\it global fits} is
guaranteed to be less than unity if an additional `intrinsic' error of
$0.21$ mag is added to the magnitude errors.

The last three points are evident from the entries in Table 1.
$N_{\rm fit}$ is the number of Cepheids accepted by R11 and $N_{\rm
  rej}$ is the number rejected. $\sigma$ is the standard deviation of
the magnitude residuals of the accepted Cepheids around the best fit
P-L relation with slope constrained to $b = -3.1$.  $\hat \chi^2$
gives the reduced $\chi^2$ computed using the magnitude errors listed
in R11. Although the dispersions exceed $0.3$ mag., the values of $\hat
\chi^2$ for many galaxies are already less than unity.  I also
list results for fits to the Wesenheit magnitudes (\ref{LMC1}).
These numbers are similar to those for the H-band fits, so adding colour 
information produces very little change to the scatter.  Since the values of $\hat \chi^2$ in
Table 1 are already low, $\hat \chi^2$ for the global fits will be
substantially less than unity if an additional $0.21$ mag. is
added in quadrature to the R11 magnitude errors. Table 1 shows that
there is simply no room for additional scatter (irrespective of colour
corrections). The choice of adding a $0.21$ magnitude error to the
Cepheids accepted by the R11 rejection algorithm is not supported by
the data.

\begin{table*}

\begin{center}
 Global fits: NGC 4258 anchor
\begin{tabular}{llllllllll} \hline
               &            \multicolumn{7}{c}{T=2.5 Rejection} & \multicolumn{2}{c}{Priors} \\
Fit        & $N_{\rm fit}$ & $\hat \chi^2_{\rm WFC3}$ & $H_0$ & $p_W$ & $b_W$ & $Z_w$ & $\sigma_{\rm int}$& $Z_w$& $b_W$   \\ \hline
1    & $485$ & $1.00$ & 71.1 (3.2) (2.4)  & 26.30 (0.17)  & -3.05 (0.13)  & -0.45 (0.15)  & 0.32  &  N & N  \\
2    & $484$ & $1.00$ & 70.3 (3.2) (2.4)  & 26.35 (0.17)  & -3.08 (0.13)  & -0.31 (0.13)  & 0.32  &  W & N  \\
3    & $481$ & $1.00$ & 69.7 (3.1) (2.3)  & 26.38 (0.16)  & -3.10 (0.12)  & -0.006 (0.020)  & 0.32  &  S & N  \\
4    & $482$ & $1.00$ & 69.0 (3.0) (2.1)  & 26.49 (0.11)  & -3.18 (0.08)  & -0.006 (0.020)  & 0.32  &  S & Y  \\ \hline

               &            \multicolumn{7}{c}{T=2.25 Rejection} & \multicolumn{2}{c}{Priors} \\
Fit        & $N_{\rm fit}$ & $\hat \chi^2_{\rm WFC3}$ & $H_0$ & $p_W$ & $b_W$ & $Z_w$ & $\sigma_{\rm int}$& $Z_w$& $b_W$   \\ \hline
5    & $458$ & $1.00$ & 70.6 (3.0) (2.1)  & 26.59 (0.15)  & -3.24 (0.11)  & -0.53 (0.13)  & 0.21  &  N & N  \\
6    & $459$ & $1.00$ & 70.3 (3.0) (2.1)  & 26.59 (0.15)  & -3.23 (0.11)  & -0.40 (0.11)  & 0.22  &  W & N  \\
7    & $447$ & $1.00$ & 70.8 (3.0) (2.0)  & 26.61 (0.14)  & -3.22 (0.10)  & -0.007 (0.020)  & 0.18  &  S & N  \\
8    & $447$ & $1.00$ & 70.8 (2.9) (2.0)  & 26.61 (0.10)  & -3.23 (0.07)  & -0.007 (0.020)  & 0.18  &  S & Y  \\ \hline

               &            \multicolumn{7}{c}{R11 Rejection} & \multicolumn{2}{c}{Priors} \\
Fit        & $N_{\rm fit}$ & $\hat \chi^2_{\rm WFC3}$ & $H_0$ & $p_W$ & $b_W$ & $Z_w$ & $\sigma_{\rm int}$& $Z_w$& $b_W$   \\ \hline
9    & $390$ & $0.64$ & 72.3 (2.8) (1.8)  & 26.43 (0.13)  & -3.10 (0.09)  & -0.33 (0.11)  & $0.21$  &  N & N  \\
10    & $390$ & $0.64$ & 72.1 (2.8) (1.8)  & 26.44 (0.13)  & -3.11 (0.10)  & -0.25 (0.10)  & $0.21$  &  W & N  \\
11    & $390$ & $0.65$ & 71.2 (2.8) (1.8)  & 26.48 (0.13)  & -3.14 (0.09)  & -0.007 (0.016)  & $0.21$  &  S & N  \\
12    & $390$ & $0.65$ & 70.8 (2.7) (1.7)  & 26.56 (0.09)  & -3.19 (0.06)  & -0.007 (0.016)  & $0.21$  &  S & Y  \\ \hline

\end{tabular}
\caption{$N_{\rm fit}$ gives the number of Cepheids accepted by the outlier rejection criteria. The
numbers in brackets give the $1\sigma$ errors on the parameters computed from the diagonals
of the inverse covariance matrix. For $H_0$ the first number in  brackets gives the total error
on the Hubble constant, including the megamaser distance error and the 
 SNe magnitude errors.  The second number in brackets
lists the error in $H_0$ from the P-L relation only. The column labelled $\sigma_{\rm int}$ lists the internal scatter.
The last two columns indicate the prior applied
to metallicity dependence $Z_W$ (N: no prior; W: weak prior; S: strong prior) and to the P-L slope $b_W$
(N: no prior; Y: prior) as summarized in (\ref{Priora}) - (\ref{Priorc}). Fits 1-4 list results for  $T=2.5$
outlier rejection,  fits 5-8 for  $T=2.25$ outlier rejection and fits 9-12 for the R11 rejection algorithm. }
\end{center}
\end{table*}

  Instead of rejecting Cepheids on a galaxy-by-galaxy basis, I reject outliers
from the global fit. As in R11, we write the P-L relation for galaxy $i$ as
\begin{equation}
m^P_{W,i} = (\mu_{0,i} - \mu_{0,4258})  + p_W + Z_W \Delta {\rm log \ (O/H)} + b_W \ {\rm log} P, \label{GPE1}
\end{equation}
and minimise,
\begin{equation}
\chi^2_{\rm WFC3} = \sum_{ij} { (m_{W,ij} - m^P_{W, i})^2 \over (\sigma^2_{e,ij} + \sigma^2_{\rm int})}, \label{GPE2}
\end{equation}
(where $j$ is the index of the Cepheid belonging to galaxy $i$ and $\sigma_{e, ij}$ is the magnitude
error listed in Table 2 of R11)  with respect to the parameters of the global fit. I use only Cepheids with periods $P < 60$ days in these fits. (See the Appendix A for remarks on the effects of extending the period range).

I set $\sigma_{\rm int}=0.30$ initially and reject Cepheids with
absolute magnitude residuals relative to the global fit that are
greater than $T \sqrt{(\sigma^2_{e, ij} + \sigma^2_{\rm int})}$ for a
chosen threshold $T$.  I then recompute $\sigma^2_{\rm int}$ to give
$\hat \chi^2 = 1$ and repeat until the fits and rejection conditions
converge. The algorithm is statistically self consistent\footnote{For
  a Gaussian distribution, the application of a threshold rejection
  will bias $\hat \chi^2$ low by factors of $0.921$ and $0.856$ for
  $T=2.5$ and $T=2.25$ respectively. These biases are neglected in
  this analysis.}, in the sense that the solutions coverge with $\hat
\chi^2=1$ for a positive value of $\sigma^2_{\rm int}$, as long as $T$
is chosen to be greater than $T=2.1$. Below I will show results for
$T=2.5$ and $T=2.25$. 
Both this and the R11 rejection algorithm are
symmetrical about the best fits and could introduce biases if the
residuals are distributed asymmetrically. Unfortunately, the Cepheid
sample is not large enough to apply statistically meaningful tests for
asymmetric residuals.

The term $\sigma_{\rm int}$ is expected 
to be non-zero since there will be scatter in the P-L relation from
the finite width of the instability strip. The analysis of the LMC Cepheids
in Section 2 suggests that at H-band the internal scatter of the P-L
relation is about $0.1$ mag (though the geometrical distribution of the LMC
Cepheids contributes to some of this scatter). R11 make no phase corrections to the H-band
magnitudes and  estimate that this introduces an additional scatter of
about $0.1$ mag. Combining these contributions suggests a minimum
value of $\sigma_{\rm int} \approx 0.14$ mag. Systematic under-estimation of magnitude
errors ({\it e.g.} crowding corrections) or contamination by outliers will
result in higher values of $\sigma_{\rm int}$.

Results of the global fits, propagated through to values of $H_0$, are
listed in Table 2. Here I have used the new NGC 4258 maser distance of
$7.60 \pm 0.23 \ {\rm Mpc}$
(H13) which improves on the maser distance of Herrnstein \etals (1999) and is
higher than the distance adopted by R11 of $7.28 \pm 0.22 \ {\rm Mpc}$ (see Riess
\etal, 2012). This change alone revises $H_0$ downwards by
approximately $3 \ \kmsMpc$.
The SNe magnitudes and errors as listed in Table 3 of
R11. Figure 2 shows the magnitude residuals with respect to  global
fit number 5 for Cepheids in NGC 4258 (left) and for the SNe host
galaxies (right). The results of Table 2 are in agreement with the
conclusion of R11, namely that the primary sensitivity to outliers
comes from a small number of highly deviant points that are easy to
identify and that are rejected by all three rejection
conditions. However, the effects of applying different rejection
criteria is non-negligible. Comparing pairs of fits 
with the different outlier rejection conditions explored in in Table 2
shows differences in $H_0$ of $\sim 1.5 \ \kmsMpc$. An estimate of the error
associated with outlier rejection should therefore be  folded into
the total error on $H_0$. (In fact, R11 add $0.8\ \kms$ in quadrature to their
final error estimate on $H_0$ to account for systematic errors such as sensitivity to
outlier rejection.)

\begin{figure*}


\includegraphics[width=85mm, angle=0]{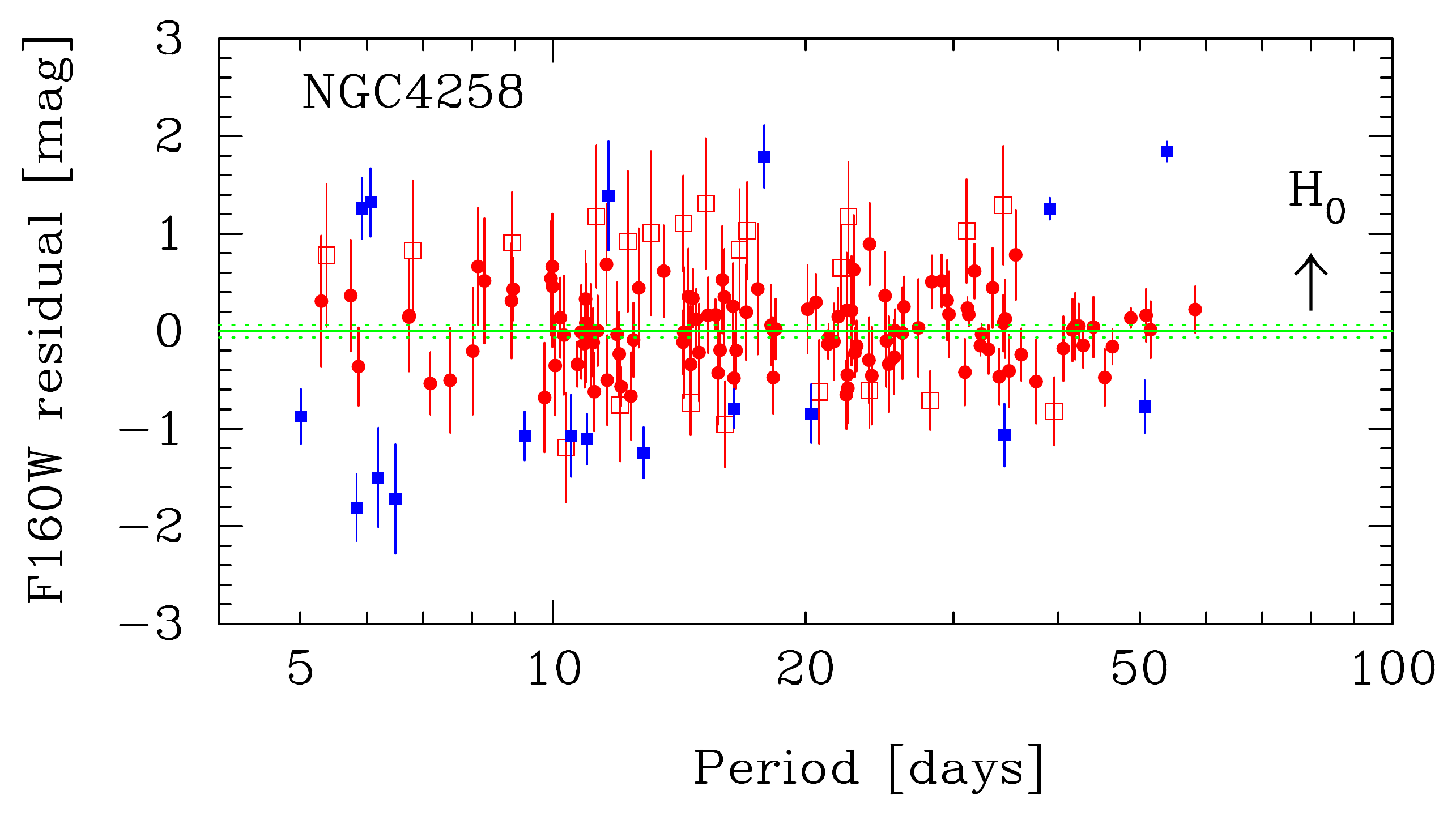}
\includegraphics[width=85mm, angle=0]{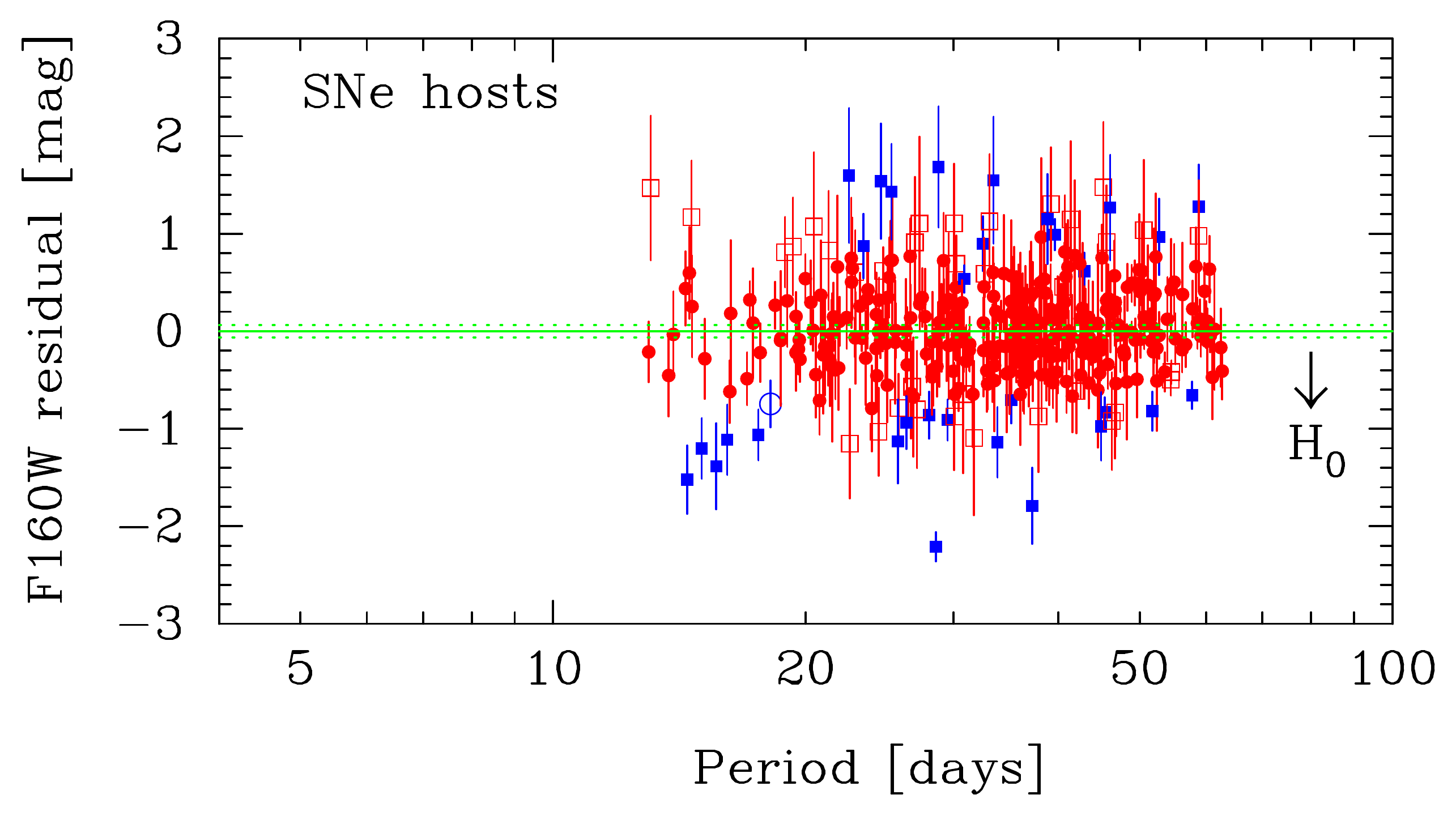}
\caption
{P-L magnitude residuals relative to global fit 5 of Table 2:
  Residuals for NCG4258 Cepheids are shown on the left and the SNe
  host galaxy Cepheids are shown on the right. Filled (red) dots  show
  residuals for Cepheids that are accepted by both the R11 and the
  $T=2.25$ rejection criterion. Filled (blue) squares  are rejected by
  both algorithms. Open (red) squares  are rejected by R11 but accepted by
  the $T=2.25$ rejection criterion and the single open (blue) circle is
  rejected by the $T=2.25$ criterion but accepted by R11. The errors
  show the H-band magnitude errors as listed in R11. The dotted lines
  show offsets that would produce changes of $\pm 2 \ \kmsMpc$ in the
  value of $H_0$ (increasing $H_0$ is shown by the direction of the
  arrow in each plot).}

\end{figure*}

\subsection{Metallicity dependence}

More worryingly, fits 1, 5 and 9 all show a strong metallicity
dependence, apparently at the $3-4\sigma$ significance level. The
metallicity dependence of the P-L relation has been controversial for
many years. At optical wavelengths, there is evidence for a
metallicity dependence of $\sim -0.25 \ {\rm mag.} \ {\rm dex}^{-1}$
(Kennicutt \etals 1998; Sakai \etals 2004, Macri \etals 2006;
Scowcroft \etals 2009). There are also theoretical arguments
(McGonegal \etals 1982) and some empirical constraints (Freedman and
Madore 2011; Scowcroft \etals 2013) to suggest that the metallicity
dependence of the P-L relation at near-IR and mid-IR wavelengths
should be weaker than at optical wavelengths. In fact, in a recent
study of the P-L relation for LMC Cepheids with spectroscopic [Fe/H]
measurements, Freedman and Madore (2011) find
\begin{equation}
Z_H = 0.05 \pm 0.02  \ {\rm mag.} \ {\rm dex}^{-1},  \label{Met1}
\end{equation}
for the metallicity dependence at H-band. An updated version of this
analysis is presented in Freedman \etals (2011) in which the
metallicity dependence of LMC Cepheids at H-band appears to be even
weaker than in equation (7). Equation (7) clearly conflicts with the
results of Table 2\footnote{The constraint of equation (7) is derived
  from a small sample of 21 LMC Cepheids. Since the metallicity
  dependence is a small effect, it is possible that subtle effects
  such as a correlation between effective temperature and metallicity,
  or between metallicty and age lead to biases in the inferred
  metallicity dependence of the P-L relation. Equation (7) is
  consistent with a weak metallicity dependence of the P-L relation at
  near-infrared wavelengths, however, both the variation of the
  metallicity dependence with waveband reported by Freedman and Madore
  (2011; see also Romaniello \etal, 2008) and their error estimates
  should be treated with caution.}.

Yet as can be seen from Figure 3, the R11 data contain sub-luminous
low metallicity Cepheids. In these plots, we show the magnitude
residual with respect to fit 5, but setting $Z_w=0$ ({\it i.e.}
neglecting the metallicity dependence of the P-L relation).  The right
hand panel of Figure 3 shows low metallicity Cepheids, almost all of
which lie below the mean relation. Some of these Cepheids are 
rejected by R11 but accepted by the $T=2.25$ rejection (open red
symbols). This is mainly because these Cepheids fail the R11 
$0.75$ mag. cut which is applied to the H-band magnitudes {\it before}
fitting for a metallicity dependence of the P-L relation.  This
difference in rejection explains why the metallicity dependence is
stronger in the $T=2.5$ and $T=2.25$ fits compared to the R11 fits.
However, the metallicity dependence is strong even with the R11
rejection algorithm. It is also worth noting that the metallicity trend extends
to Cepheids with $12 + {\rm log(O/H)} > 8.6$. Eliminating the small number of stars
with $12 + {\rm log(O/H)} < 8.6$ reduces, but does not eliminate, the metallicity
dependence of the global fits.

The strong metallicity dependence in the R11 sample may 
indicate an unidentified systematic in the data. For example, 
the low metallicity systems may be blended Type II Cepheids 
that are not identified by the outlier rejection algorithm. Another
possibility might be errors in the crowding bias corrections (since the low
metallicity Cepheids have lower than average crowding bias
corrections). The effect is so large that it seems unlikely that
the discrepancy between (\ref{Met1}) and the R11 data is related to
the use of nebular [O/H] estimates of metallicity rather than
spectroscopic [Fe/H] metallicities of individual Cepheids.
Additional constraints on  $Z_W$, based on [O/H] metallicities, are discussed
in Section 5.

\begin{figure*}


\includegraphics[width=85mm, angle=0]{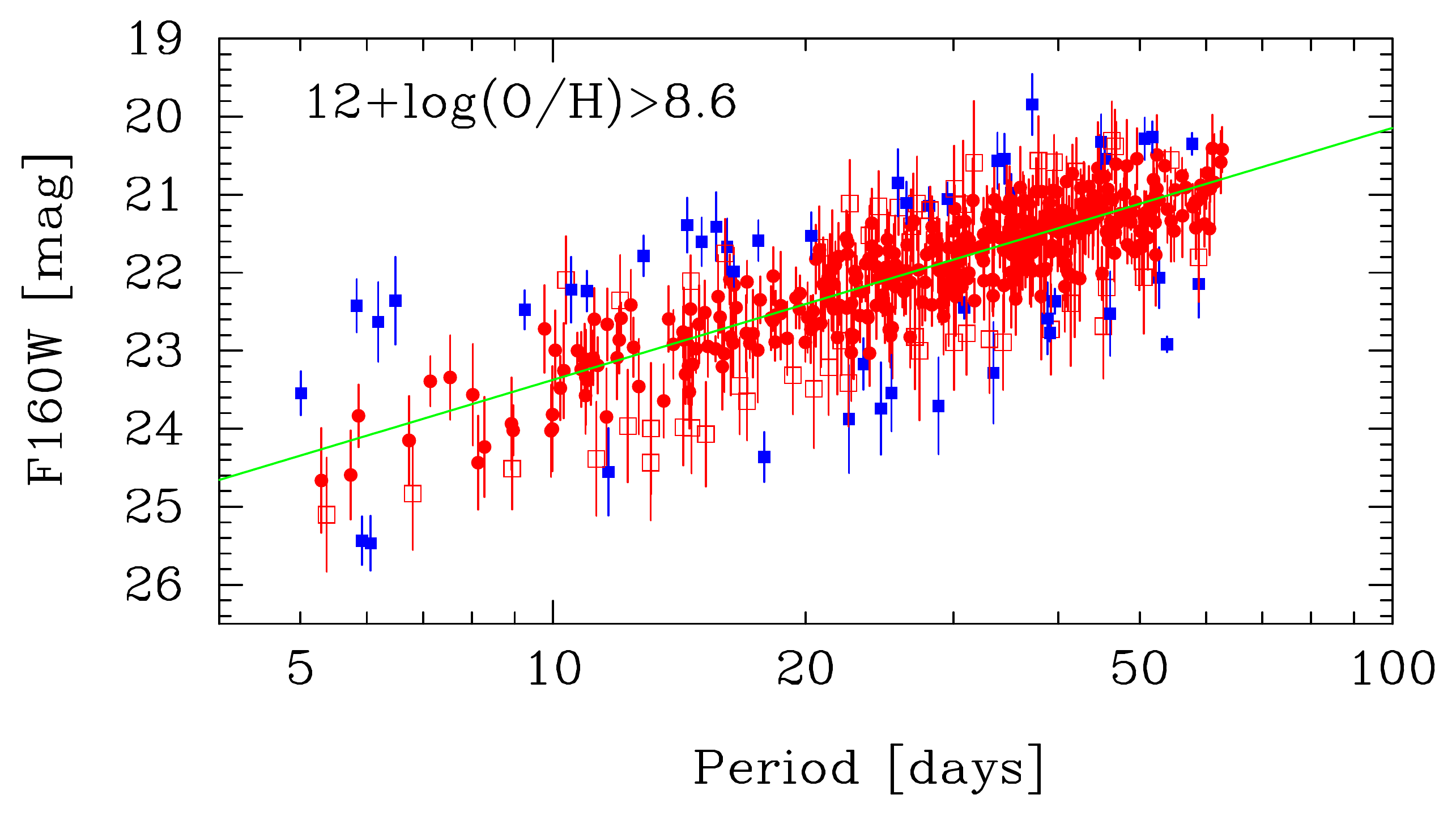}
\includegraphics[width=85mm, angle=0]{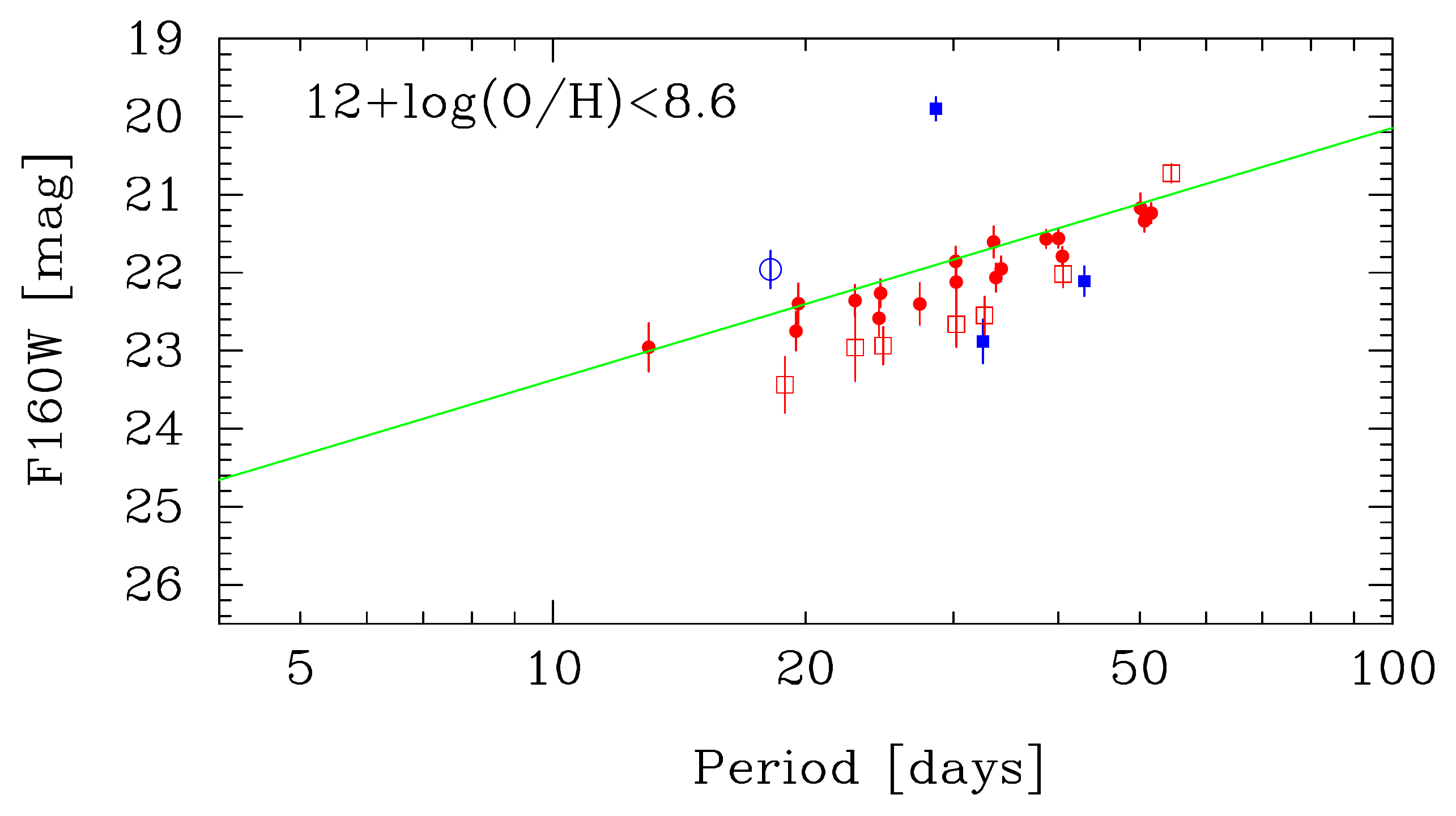}
\caption
{P-L relations for the data used in fit 5. The symbols and colour coding of the points are as in 
Figure 3. In these plots, the F160W magnitudes have not been corrected
 for a metallicity dependence. The figure to the left shows high metallicity Cepheids
and the figure to the right shows low metallicity Cepheids. The lines show the best fit P-L
relation for the entire sample.}
\end{figure*}

Another way of reducing the sensitivity to outliers and possible systematics
in the data is to impose priors on the parameters of the P-L relation.
I have explored the following priors:
\beglet
\begin{eqnarray}
\langle Z_w \rangle = 0,  &   \sigma_{Z_w} = 0.25, & \quad {\rm weak \ metallicity \ prior},  \label{Priora}\\
\langle Z_w \rangle = 0,  &   \sigma_{Z_w} = 0.02, &\quad {\rm strong \ metallicity\  prior}, \label{Priorb}\\
\langle b \rangle  = -3.23,  &   \sigma_{b} = 0.10, & \quad {\rm P-L \ slope \ prior}  \label{Priorc}.
\end{eqnarray}
\endlet
The first of these imposes a weak prior on the metallicity dependence of
the P-L relation. The second imposes a strong prior, effectively eliminating
a  metallicity dependence of the P-L relation. The last condition is
 motivated by constraints on the slope of the LMC P-L relation discussed in Section 2.

The results of applying these priors are listed in Table 2. For the R11 
rejection, applying these priors drives $H_0$  downwards by $\sim 1.5 \ \kmsMpc$. 
The results for the $T=2.25$ rejection are, however, extremely stable
to the imposition of the priors.

To determine a final `best-estimate' of $H_0$ using NGC 4258 as an
anchor, I have averaged the $H_0$ values of fits 3, 7, 11, and added
the scatter between these estimates in quadrature with the largest of the error estimates. I
have also added a $1 \ \kmsMpc$ error to account for systematics
associated with the SNe magnitudes and light curve fitting.  
This gives
\begin{equation}
H_0 = 70.6  \pm  3.3 \ \kmsMpc,   \quad {\rm NGC \ 4258}.  \label{H04258}
\end{equation}

This value is lower than the value $H_0 = 72 \pm 3 \ \kmsMpc$
quoted by H13 using the revised maser distance to NGC 4258. This difference
is caused mainly by the use of difference outlier rejection criteria and the imposition
of a  metallicity prior. Note also that (\ref{H04258}) is within $1\sigma$
of the \Planck\ value of $H_0$ for the base \lcdm\ model.

\medskip

\section{Using the LMC and Milky Way Cepheids as anchors}

\begin{table*}

\begin{center}

\medskip

           Global fits: LMC anchor

\begin{tabular}{llllllllll} \hline
               &            \multicolumn{7}{c}{T=2.5 Rejection} & \multicolumn{2}{c}{Priors} \\
Fit        & $N_{\rm fit}$ & $\hat \chi^2_{\rm WFC3}$ & $H_0$ & $p_W$ & $b_W$ & $Z_w$ & $\sigma_{\rm int}$& $Z_w$& $b_W$   \\ \hline
13    & $479$ & $1.00$ & 68.6 (3.5) (2.2)  & 15.59 (0.08)  & -3.23 (0.05)  & -0.47 (0.14)  & 0.32  &  N & N  \\
14    & $478$ & $1.00$ & 69.7 (3.4) (2.1)  & 15.64 (0.08)  & -3.23 (0.05)  & -0.35 (0.13)  & 0.32  &  W & N  \\
15    & $481$ & $1.00$ & 73.5 (2.9) (1.9)  & 15.76 (0.06)  & -3.21 (0.05)  & -0.005 (0.020)  & 0.32  &  S & N  \\
16    & $480$ & $1.00$ & 73.5 (2.8) (1.9)  & 15.77 (0.06)  & -3.22 (0.05)  & -0.006 (0.020)  & 0.32  &  S & Y  \\ \hline

               &            \multicolumn{7}{c}{T=2.25 Rejection} & \multicolumn{2}{c}{Priors} \\
Fit        & $N_{\rm fit}$ & $\hat \chi^2_{\rm WFC3}$ & $H_0$ & $p_W$ & $b_W$ & $Z_w$ & $\sigma_{\rm int}$& $Z_w$& $b_W$   \\ \hline
17    & $458$ & $1.00$ & 67.4 (3.2) (2.0)  & 15.57 (0.08)  & -3.23 (0.05)  & -0.53 (0.13)  & 0.21  &  N & N  \\
18    & $459$ & $1.00$ & 68.7 (3.2) (2.0)  & 15.62 (0.08)  & -3.23 (0.05)  & -0.40 (0.11)  & 0.22  &  W & N  \\
19    & $447$ & $1.00$ & 73.3 (2.8) (1.8)  & 15.78 (0.06)  & -3.23 (0.05)  & -0.007 (0.020)  & 0.18  &  S & N  \\
20    & $447$ & $1.00$ & 73.3 (2.8) (1.8)  & 15.78 (0.05)  & -3.23 (0.05)  & -0.007 (0.020)  & 0.18  &  S & Y  \\ \hline

               &            \multicolumn{7}{c}{R11 Rejection} & \multicolumn{2}{c}{Priors} \\
Fit        & $N_{\rm fit}$ & $\hat \chi^2_{\rm WFC3}$ & $H_0$ & $p_W$ & $b_W$ & $Z_w$ & $\sigma_{\rm int}$& $Z_w$& $b_W$   \\ \hline
21    & $390$ & $0.64$ & 70.7 (3.1) (1.8)  & 15.63 (0.07)  & -3.21 (0.04)  & -0.31  (0.11)  & $0.21$  &  N & N  \\
22    & $390$ & $0.64$ & 71.3 (3.0) (1.8)  & 15.64 (0.06)  & -3.21 (0.04)  & -0.24 (0.10)  & $0.21$  &  W & N  \\
23    & $390$ & $0.65$ & 73.5 (2.7) (1.6)  & 15.76 (0.05)  & -3.21 (0.04)  & -0.006 (0.016)  & $0.21$  &  S & N  \\
24    & $390$ & $0.65$ & 73.4 (2.6) (1.5)  & 15.77 (0.05)  & -3.22 (0.04)  & -0.006 (0.016)  & $0.21$  &  S & Y  \\ \hline

\end{tabular}

\medskip

             Global fits: MW Cepheids anchor
\begin{tabular}{llllllllll} \hline
               &            \multicolumn{7}{c}{T=2.5 Rejection} & \multicolumn{2}{c}{Priors} \\
Fit        & $N_{\rm fit}$ & $\hat \chi^2_{\rm WFC3}$ & $H_0$ & $M_W$ & $b_W$ & $Z_w$ & $\sigma_{\rm int}$& $Z_w$& $b_W$   \\ \hline
25    & $486$ & $1.00$ & 76.5 (4.0) (3.7)  & -5.89 (0.05)  & -3.22 (0.11)  & -0.47 (0.15)  & 0.30  &  N & N  \\
26    & $484$ & $1.00$ & 77.7 (4.1) (3.8)  & -5.88 (0.05)  & -3.15 (0.10)  & -0.31 (0.13)  & 0.32  &  W & N  \\
27    & $482$ & $1.00$ & 76.5 (3.9) (3.6)  & -5.89 (0.05)  & -3.17 (0.11)  & -0.006 (0.020)  & 0.32  &  S & N  \\
28    & $482$ & $1.00$ & 75.7 (3.4) (3.1)  & -5.89 (0.05)  & -3.20 (0.07)  & -0.006 (0.020)  & 0.32  &  S & Y  \\ \hline

               &            \multicolumn{7}{c}{T=2.25 Rejection} & \multicolumn{2}{c}{Priors} \\
Fit        & $N_{\rm fit}$ & $\hat \chi^2_{\rm WFC3}$ & $H_0$ & $M_W$ & $b_W$ & $Z_w$ & $\sigma_{\rm int}$& $Z_w$& $b_W$   \\ \hline
29    & $458$ & $1.00$ & 75.2 (3.8) (3.5)  & -5.90 (0.05)  & -3.26 (0.10)  & -0.53 (0.13)  & 0.21  &  N & N  \\
30    & $459$ & $1.00$ & 74.8 (3.7) (3.5)  & -5.90 (0.05)  & -3.26 (0.10)  & -0.40 (0.11)  & 0.22  &  W & N  \\
31    & $459$ & $1.00$ & 73.6 (3.6) (3.3)  & -5.90 (0.05)  & -3.26 (0.10)  & -0.009 (0.020)  & 0.23  &  S & N  \\
32    & $447$ & $1.00$ & 74.7 (3.2) (2.9)  & -5.90 (0.05)  & -3.24 (0.07)  & -0.007 (0.020)  & 0.18  &  S & Y  \\ \hline

               &            \multicolumn{7}{c}{R11 Rejection} & \multicolumn{2}{c}{Priors} \\
Fit        & $N_{\rm fit}$ & $\hat \chi^2_{\rm WFC3}$ & $H_0$ & $M_W$ & $b_W$ & $Z_w$ & $\sigma_{\rm int}$& $Z_w$& $b_W$   \\ \hline
33    & $390$ & $0.64$ & 77.9 (3.3) (2.9)  & -5.88 (0.04)  & -3.15 (0.08)  & -0.32 (0.11)  & $0.21$  &  N & N  \\
34    & $390$ & $0.64$ & 77.4 (3.3) (2.9)  & -5.88 (0.04)  & -3.16 (0.08)  & -0.24 (0.10)  & $0.21$  &  W & N  \\
35    & $390$ & $0.65$ & 76.0 (3.2) (2.8)  & -5.89 (0.04)  & -3.18 (0.08)  & -0.006 (0.016)  & $0.21$  &  S & N  \\
36    & $390$ & $0.65$ & 75.4 (2.8) (2.4)  & -5.89 (0.04)  & -3.21 (0.06)  & -0.006 (0.016)  & $0.21$  &  S & Y  \\ \hline

\end{tabular}
\end{center}

\smallskip

\caption{
As in Table 2, the numbers in brackets give the $1\sigma$ errors on the parameters computed from the diagonals
of the inverse covariance matrix. For $H_0$ the first number in  brackets adds the errors arising from the
distance anchors and SNe magnitudes. The second number in brackets
lists the error in $H_0$ from the P-L relation alone. $N_{\rm fit}$ gives the number of R11 Cepheids
retained in the fits. The remaining columns are as defined in Table 3.}

\end{table*}

\subsection{The LMC Cepheids}

Since the mean metallicities of NGC 4258 and SNe hosts are similar
($12 + {\rm log\ (O/H)} \approx 8.9$), the metallicity dependence
discussed in Section 3.2 has a small but non-negligible effect on $H_0$ if NGC
4258 is used  as a distance anchor. However, if we use the LMC as an anchor
(for which we assume $12 +{\rm log \ (O/H)} = 8.5$), the metallicity
dependence of the P-L relation becomes significant. Values as large as
$Z_W \sim -0.3 \ {\rm mag} \ {\rm dex}^{-1} $ lead to a substantial
reduction in the value of $H_0$ compared to fits in which $Z_W$ is
constrained to be zero (see Table 3).

For the LMC distance, I use the new eclipsing binary distance of
$49.97 \pm 1.13 \ {\rm kpc}$ determined by Pietrzy\'nski \etals
(2013). I minimise the sum of the $\chi^2$ of equations (\ref{LMC3})
and (\ref{GPE2}) (which are denoted $\chi^2_{\rm LMC}$ and
$\chi^2_{\rm WFC3}$ respectively) with respect to the parameters of
the P-L relation applying the rejection criteria described in Section
3.1 to the R11 Cepheids.  The LMC Cepheid sample is as described in
Section 2 and $\sigma_{\rm int, LMC}$ is kept fixed at $0.113$. In
applying the $T=2.5$ and $T=2.25$ rejection criteria, $\sigma_{\rm
  int, WFC3}$ is adjusted to maintain $\hat \chi^2_{\rm WFC3} = 1$. In
using the LMC (or MW) Cepheids as a distance anchor, the only role of
the NGC 4258 Cepheids in determining $H_0$ is to influence the slope
and metallicity dependence of the global fit to the P-L relation. I
have chosen to retain the NGC 4258 Cepheids (as did R11), though the 
results are very similar if these Cepheids are excluded.

Table 3 lists the results for the three rejection criteria. If no
priors are included, the values of $H_0$ show some sensitivity to the
rejection algorithm. This sensitivity is caused by the low metallicity
Cepheids in the R11 sample, which pull $Z_W$ towards negative values
(less strongly for the R11 rejection algorithm). Applying the strong
metallicity prior, we find very little sensitivity to the rejection
algorithm. Furthermore, imposing the slope prior of (\ref{Priorc}) has
very little effect on the solutions because the slopes of the global
fits are well constrained by the LMC Cepheids. Averaging the results
for fits 15, 19 and 23, we find:
\begin{equation}
H_0 = 73.4  \pm  3.1 \ \kmsMpc,   \qquad {\rm LMC, \ strong \ metallicity \ prior}.  \label{H0LMC}
\end{equation}
More generally, the solutions of Table 3 show a strong sensitivity to the 
metallicity dependence of the P-L relation:
\begin{equation}
H_0  \approx (73.4 + 10 Z_w)  \ \kmsMpc.
\end{equation}
An accurate determination of $H_0$ using the LMC as an anchor
therefore requires a precise determination of $Z_w$.  As discussed in
Section 3.2, the strong metallicity dependence at H-band wavelengths
derived from the R11 data conflict with the weak metallicity
dependence of equation (7).  Applying the strong metallicity prior
(\ref{H0LMC}),  $H_0$ is in tension, at about the $1.9\sigma$, level
with the \Planck\ base \LCDM\ value for $H_0$. This tension can be relieved
if the metallicity dependence of the P-L relation is somewhat stronger than
implied by equation (7).

\subsection{The MW Cepheids}

I use the sample of $13$ MW Cepheids with parallax measurements and
photometry as listed in van Leeuwen \etals (2007) (eliminating
Polaris).  The Wesenheit P-L relation (HVI photometry) for these
Cepheids is shown in Figure 5. A fit to these gives
\begin{equation}
M_W= -5.91 \pm 0.17, \quad b_W = -3.29 \pm 0.17, \quad \hat \chi^2=0.57,  \label{MW1}
\end{equation}
where $M_W$ replaces $A$ in equation (1).
Note that the slope is not well constrained because of the dearth of Cepheids with periods
greater than 10 days in this small sample. The slope is, however, 
compatible with the slope determined from the
LMC Cepheids. The reduced $\chi^2$ of this sample is smaller than unity, but because of the
small sample size this is not statistically significant and leaves room for a significant
`internal dispersion' (which cannot be well constrained from $\hat \chi^2$). In the fits below, 
I adopt an internal dispersion of $\sigma_{int} = 0.10$,
consistent with the internal dispersion of the LMC sample. The parameters of the MW Cepheid
fits discussed below are insensitive to this value, though adopting $\sigma_{int}=0.10$ has
the effect of slightly downweighting the MW Cepheids compared to the LMC and/or NGC 4258 when
combining  distance anchors.

The global fits using the MW parallax distances are summarized in Table 3. The metallicity
dependences of these fits are skewed by the low metallicity outliers in the R11 sample. These
raise $H_0$ and introduce a sensitivity to the outlier rejection algorithm. Imposing the
strong metallicity and P-L slope priors reduces the sensitivity to these outliers.
Averaging the results of fits 28, 32 and 36 gives
\begin{equation}
H_0 = 75.3  \pm  3.5 \ \kmsMpc,   \qquad {\rm MW, \ strong \ metallicity \ and  \ slope \ priors}.  \label{H0MW}
\end{equation}
This result is in tension at about the $2.1\sigma$, level with the
\Planck\ base \LCDM\ value for $H_0$. Because of the small size of the
MW Cepheid sample, the instability strip is not well sampled. In
addition, the lack of overlap between the periods of the MW Cepheids
and the R11 sample leads to a sensitivity of $H_0$ to the slope prior
and to the choice of period range for the R11 Cepheids (see Appendix
A).  Coupled with possible systematic errors associated with matching
ground-based and HST photometry, I consider the MW Cepheids to be the
least reliable of the three distance anchors.

\begin{figure}

\begin{center}
\includegraphics[width=120mm, angle=0]{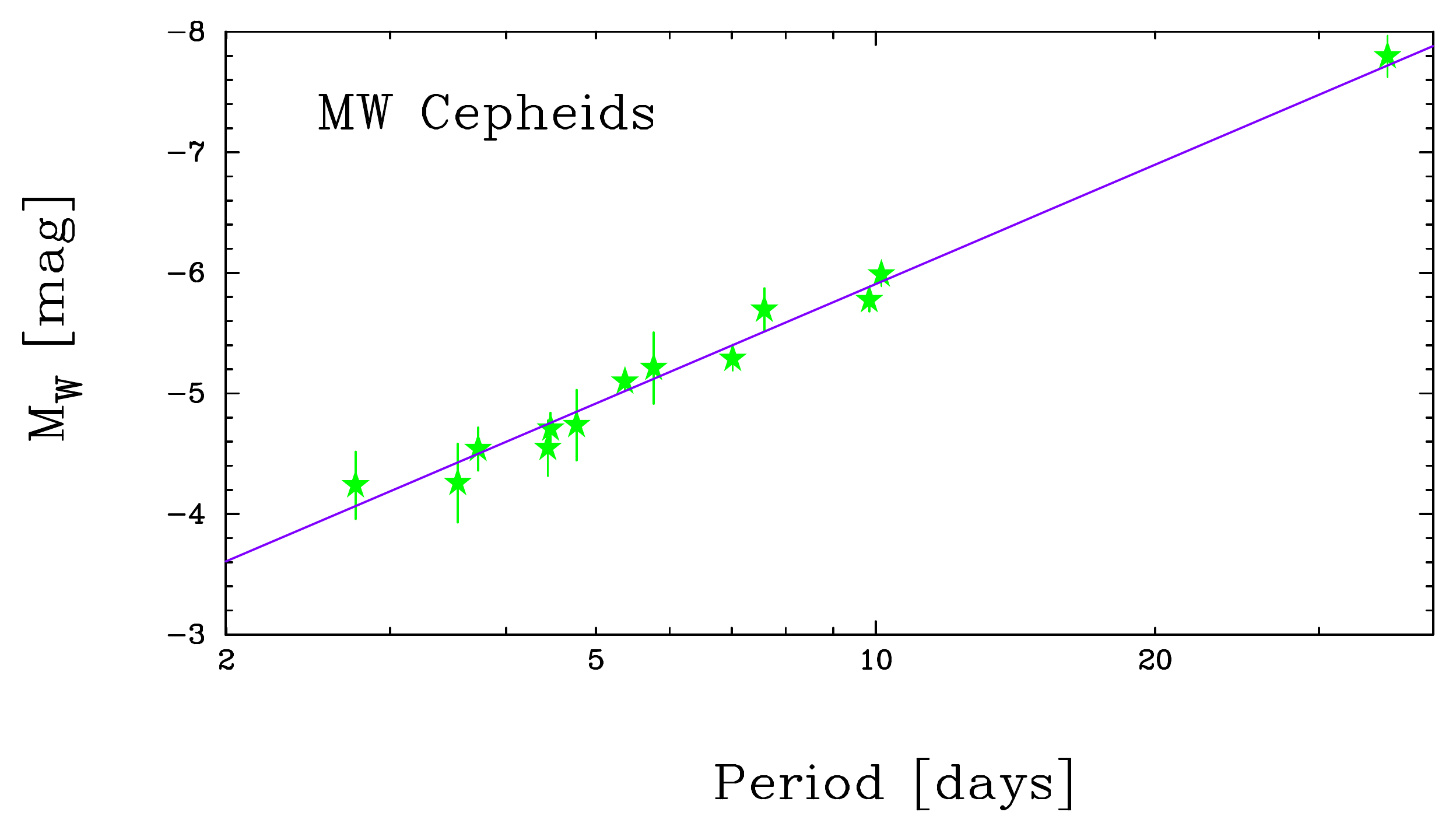}
\end{center}
\caption
{Period-luminosity relation for  13 MW Cepheids with parallax measurements
(from van Leeuwen \etals 2007). The line shows the best
fit of (\ref{MW1})}. 
\end{figure}

\section{Combining Distance Anchors}

\begin{table*}

\begin{center}

\medskip

              \center{Global fits: NGC 4258+LMC  anchors \hfill}
{\scriptsize
\begin{tabular}{llllllllll} \hline
               &            \multicolumn{7}{c}{T=2.5 Rejection} & \multicolumn{2}{c}{Priors} \\
Fit        & $N_{\rm fit}$ & $\hat \chi^2_{\rm WFC3}$ & $H_0$ & $M_W$ & $b_W$ & $Z_w$ & $\sigma_{\rm int}$& $Z_w$& $b_W$   \\ \hline
37    & $477$ & $1.00$ & 69.3 (2.4)   & -6.10 (0.06)  & -3.22 (0.05)  & -0.45 (0.13)  & 0.30  &  N & N  \\
38    & $478$ & $1.00$ & 69.8 (2.4)   & -6.07 (0.06)  & -3.22 (0.05)  & -0.35 (0.11)  & 0.30  &  W & N  \\
39    & $477$ & $1.00$ & 71.4 (2.4)   & -5.99 (0.05)  & -3.23 (0.05)  & -0.009 (0.020)  & 0.30  &  S & N  \\
40    & $477$ & $1.00$ & 71.5 (2.4)   & -5.99 (0.05)  & -3.24 (0.05)  & -0.009 (0.020)  & 0.30  &  S & Y  \\ \hline

               &            \multicolumn{7}{c}{T=2.25 Rejection} & \multicolumn{2}{c}{Priors} \\
Fit        & $N_{\rm fit}$ & $\hat \chi^2_{\rm WFC3}$ & $H_0$ & $M_W$ & $b_W$ & $Z_w$ & $\sigma_{\rm int}$& $Z_w$& $b_W$   \\ \hline
41    & $458$ & $1.00$ & 69.0 (2.3)   & -6.10 (0.06)  & -3.23 (0.05)  & -0.48 (0.12)  & 0.21  &  N & N  \\
42    & $459$ & $1.00$ & 69.4 (2.3)   & -6.07 (0.06)  & -3.23 (0.05)  & -0.38 (0.11)  & 0.22  &  W & N  \\
43    & $448$ & $1.00$ & 71.8 (2.3)   & -5.97 (0.05)  & -3.24 (0.05)  & -0.009 (0.020)  & 0.18  &  S & N  \\
44    & $448$ & $1.00$ & 71.9 (2.3)   & -5.97 (0.05)  & -3.23 (0.04)  & -0.009 (0.020)  & 0.18  &  S & Y  \\ \hline

               &            \multicolumn{7}{c}{R11 Rejection} & \multicolumn{2}{c}{Priors} \\
Fit        & $N_{\rm fit}$ & $\hat \chi^2_{\rm WFC3}$ & $H_0$ & $M_W$ & $b_W$ & $Z_w$ & $\sigma_{\rm int}$& $Z_w$& $b_W$   \\ \hline
45    & $390$ & $0.64$ & 71.0 (1.9)   & -6.06 (0.05)  & -3.21 (0.04)  & -0.29 (0.10)  & $0.21$  &  N & N  \\
46    & $390$ & $0.64$ & 71.2 (1.9)   & -6.04 (0.05)  & -3.21 (0.04)  & -0.24 (0.09)  & $0.21$  &  W & N  \\
47    & $390$ & $0.66$ & 72.1 (1.9)   & -5.98 (0.04)  & -3.22 (0.04)  & -0.008 (0.016)  & $0.21$  &  S & N  \\
48    & $390$ & $0.65$ & 72.1 (1.9)   & -5.98 (0.04)  & -3.22 (0.04)  & -0.008 (0.016)  & $0.21$  &  S & Y  \\ \hline

\end{tabular}}

\medskip

\medskip

             Global fits:  NGC 4258+MW Cepheid  anchors
{\scriptsize
\begin{tabular}{llllllllll} \hline
               &            \multicolumn{7}{c}{T=2.5 Rejection} & \multicolumn{2}{c}{Priors} \\
Fit        & $N_{\rm fit}$ & $\hat \chi^2_{\rm WFC3}$ & $H_0$ & $M_W$ & $b_W$ & $Z_w$ & $\sigma_{\rm int}$& $Z_w$& $b_W$   \\ \hline
49    & $476$ & $1.00$ & 72.2 (2.9)   & -5.96 (0.05)  & -3.31 (0.09)  & -0.51 (0.14)  & 0.29  &  N & N  \\
50    & $478$ & $1.00$ & 71.7 (2.9)   & -5.96 (0.05)  & -3.32 (0.09)  & -0.35 (0.13)  & 0.30  &  W & N  \\
51    & $477$ & $1.00$ & 70.9 (2.8)   & -5.96 (0.05)  & -3.32 (0.09)  & -0.007 (0.020)  & 0.30  &  S & N  \\
52    & $477$ & $1.00$ & 71.6 (2.6)   & -5.96 (0.05)  & -3.28 (0.07)  & -0.007 (0.020)  & 0.30  &  S & Y  \\ \hline

               &            \multicolumn{7}{c}{T=2.25 Rejection} & \multicolumn{2}{c}{Priors} \\
Fit        & $N_{\rm fit}$ & $\hat \chi^2_{\rm WFC3}$ & $H_0$ & $M_W$ & $b_W$ & $Z_w$ & $\sigma_{\rm int}$& $Z_w$& $b_W$   \\ \hline
53    & $456$ & $1.00$ & 72.1 (2.8)   & -5.95 (0.06)  & -3.32 (0.09)  & -0.51 (0.12)  & 0.21  &  N & N  \\
54    & $455$ & $1.00$ & 71.4 (2.7)   & -5.95 (0.05)  & -3.34 (0.09)  & -0.38 (0.11)  & 0.21  &  W & N  \\
55    & $447$ & $1.00$ & 71.5 (2.7)   & -5.94 (0.05)  & -3.30 (0.08)  & -0.007 (0.020)  & 0.18  &  S & N  \\
56    & $448$ & $1.00$ & 72.2 (2.5)   & -5.95 (0.05)  & -3.27 (0.05)  & -0.007 (0.020)  & 0.19  &  S & Y  \\ \hline

               &            \multicolumn{7}{c}{R11 Rejection} & \multicolumn{2}{c}{Priors} \\
Fit        & $N_{\rm fit}$ & $\hat \chi^2_{\rm WFC3}$ & $H_0$ & $M_W$ & $b_W$ & $Z_w$ & $\sigma_{\rm int}$& $Z_w$& $b_W$   \\ \hline
57    & $390$ & $0.65$ & 73.8 (2.3)   & -5.95 (0.04)  & -3.23 (0.07)  & -0.30 (0.11)  & $0.21$  &  N & N  \\
58    & $390$ & $0.64$ & 73.5 (2.3)   & -5.95 (0.04)  & -3.24 (0.07)  & -0.23 (0.07)  & $0.21$  &  W & N  \\
59    & $390$ & $0.66$ & 72.6 (2.3)   & -5.95 (0.04)  & -3.25 (0.07)  & -0.006 (0.016)  & $0.21$  &  S & N  \\
60    & $390$ & $0.66$ & 72.7 (2.1)   & -5.95 (0.04)  & -3.24 (0.05)  & -0.006 (0.016)  & $0.21$  &  S & Y  \\ \hline

\end{tabular}}

\medskip

\medskip

           Global fits:  LMC+MW Cepheid anchors

{\scriptsize
\begin{tabular}{llllllllll} \hline
               &            \multicolumn{7}{c}{T=2.5 Rejection} & \multicolumn{2}{c}{Priors} \\
Fit        & $N_{\rm fit}$ & $\hat \chi^2_{\rm WFC3}$ & $H_0$ & $M_W$ & $b_W$ & $Z_w$ & $\sigma_{\rm int}$& $Z_w$& $b_W$   \\ \hline
61    & $479$ & $1.00$ & 72.4  (2.5)  & -5.98 (0.05)  & -3.25 (0.05)  & -0.30 (0.12)  & 0.31  &  N & N  \\
62    & $479$ & $1.00$ & 72.6  (2.5)  &  -5.96 (0.05)  & -3.25 (0.05)  & -0.24 (0.11)  & 0.31  &  W & N  \\
63    & $480$ & $1.00$ & 74.1 (2.5)   & -5.92 (0.05)  & -3.23 (0.05)  & -0.005 (0.020)  & 0.32  &  S & N  \\
64    & $477$ & $1.00$ & 74.1 (2.5)   & -5.92 (0.05)  & -3.23 (0.04)  & -0.005 (0.020)  & 0.32  &  S & Y  \\ \hline

               &            \multicolumn{7}{c}{T=2.25 Rejection} & \multicolumn{2}{c}{Priors} \\
Fit        & $N_{\rm fit}$ & $\hat \chi^2_{\rm WFC3}$ & $H_0$ & $M_W$ & $b_W$ & $Z_w$ & $\sigma_{\rm int}$& $Z_w$& $b_W$   \\ \hline
65    & $459$ & $1.00$ & 71.6 (2.4)   & -5.99 (0.05)  & -3.26 (0.05)  & -0.36 (0.11)  & 0.21  &  N & N  \\
66    & $455$ & $1.00$ & 71.9 (2.4)   & -5.97 (0.05)  & -3.26 (0.05)  & -0.28 (0.10)  & 0.21  &  W & N  \\
67    & $448$ & $1.00$ & 73.6 (2.4)  &  -5.92 (0.05)  & -3.24 (0.05)  & -0.007 (0.020)  & 0.18  &  S & N  \\
68    & $448$ & $1.00$ & 73.7 (2.4)   & -5.92 (0.05)  & -3.24 (0.04)  & -0.007 (0.020)  & 0.18  &  S & Y  \\ \hline

               &            \multicolumn{7}{c}{R11 Rejection} & \multicolumn{2}{c}{Priors} \\
Fit        & $N_{\rm fit}$ & $\hat \chi^2_{\rm WFC3}$ & $H_0$ & $M_W$ & $b_W$ & $Z_w$ & $\sigma_{\rm int}$& $Z_w$& $b_W$   \\ \hline
69    & $390$ & $0.64$ & 73.4 (2.0)  & -5.96 (0.04)  & -3.23 (0.04)  & -0.20  (0.09)  & $0.21$  &  N & N  \\
70    & $390$ & $0.65$ & 73.6 (2.0)  & -5.95 (0.04)  & -3.23 (0.04)  & -0.16 (0.09)  & $0.21$  &  W & N  \\
71    & $390$ & $0.65$ & 74.1  (2.0)  & -5.92 (0.04)  & -3.23 (0.04)  & -0.005 (0.016)  & $0.21$  &  S & N  \\
72    & $390$ & $0.65$ & 74.0  (2.0)  & -5.92 (0.04)  & -3.23 (0.04)  & -0.006 (0.016)  & $0.21$  &  S & Y  \\ \hline

\end{tabular}}

\smallskip

\caption{Solutions combining distance anchors. The columns are as defined in Tables 2 and 3.}

\end{center}

\addtocounter{table}{-1}

\end{table*}

\begin{table*}

\begin{center}

           Global fits: NGC 4258+LMC+MW Cepheids anchor

{\scriptsize  
\begin{tabular}{llllllllll} \hline
               &            \multicolumn{7}{c}{T=2.5 Rejection} & \multicolumn{2}{c}{Priors} \\
Fit        & $N_{\rm fit}$ & $\hat \chi^2_{\rm WFC3}$ & $H_0$ & $M_W$ & $b_W$ & $Z_w$ & $\sigma_{\rm int}$& $Z_w$& $b_W$   \\ \hline
73    & $479$ & $1.00$ & 71.5  (2.2)  & -6.00 (0.04)  & -3.26 (0.05)  & -0.33 (0.12)  & 0.30  &  N & N  \\
74    & $479$ & $1.00$ & 71.6  (2.3)  &  -5.99 (0.04)  & -3.26 (0.05)  & -0.27 (0.11)  & 0.31  &  W & N  \\
75    & $477$ & $1.00$ & 72.3 (2.3)   & -5.95 (0.04)  & -3.26 (0.05)  & -0.008 (0.020)  & 0.30  &  S & N  \\
76    & $477$ & $1.00$ & 72.4 (2.2)   & -5.95 (0.04)  & -3.25 (0.04)  & -0.008 (0.020)  & 0.30  &  S & Y  \\ \hline

               &            \multicolumn{7}{c}{T=2.25 Rejection} & \multicolumn{2}{c}{Priors} \\
Fit        & $N_{\rm fit}$ & $\hat \chi^2_{\rm WFC3}$ & $H_0$ & $M_W$ & $b_W$ & $Z_w$ & $\sigma_{\rm int}$& $Z_w$& $b_W$   \\ \hline
77    & $455$ & $1.00$ & 71.3 (2.2)   & -5.99 (0.04)  & -3.26 (0.05)  & -0.34 (0.11)  & 0.21  &  N & N  \\
78    & $455$ & $1.00$ & 71.4 (2.2)   & -5.98 (0.04)  & -3.26 (0.05)  & -0.29 (0.10)  & 0.21  &  W & N  \\
79    & $447$ & $1.00$ & 72.4 (2.2)  &  -5.94 (0.04)  & -3.25 (0.05)  & -0.007 (0.020)  & 0.18  &  S & N  \\
80    & $447$ & $1.00$ & 72.5 (2.2)   & -5.94 (0.04)  & -3.25 (0.04)  & -0.007 (0.020)  & 0.18  &  S & Y  \\ \hline

               &            \multicolumn{7}{c}{R11 Rejection} & \multicolumn{2}{c}{Priors} \\
Fit        & $N_{\rm fit}$ & $\hat \chi^2_{\rm WFC3}$ & $H_0$ & $M_W$ & $b_W$ & $Z_w$ & $\sigma_{\rm int}$& $Z_w$& $b_W$   \\ \hline
81    & $390$ & $0.64$ & 72.6 (1.8)  & -5.98 (0.03)  & -3.24 (0.04)  & -0.22  (0.09)  & $0.21$  &  N & N  \\
82    & $390$ & $0.65$ & 72.6 (1.8)  & -5.97 (0.03)  & -3.24 (0.04)  & -0.18 (0.08)  & $0.21$  &  W & N  \\
83    & $390$ & $0.66$ & 72.9  (1.8)  & -5.95 (0.03)  & -3.24 (0.04)  & -0.006 (0.016)  & $0.21$  &  S & N  \\
84    & $390$ & $0.66$ & 72.9  (1.8)  & -5.95 (0.03)  & -3.24 (0.04)  & -0.006 (0.016)  & $0.21$  &  S & Y  \\ \hline

\end{tabular}}

\smallskip

\caption{{\bf (contd.)} Solutions combining distance anchors.}

\end{center}

\end{table*}

\medskip

\subsection{Joint solutions}

Whereas R11 found similar values of $H_0$ using NGC 4258 and the MW Cepheids as distance
anchors, with $H_0$ for the LMC lying low,   we find reasonable agreement between
the $H_0$ values for the LMC and MW Cepheids (equations 10 and 13) with $H_0$ for 
NGC 4258 lying low (equation 9). There are two reasons why our results differ from 
R11. Firstly, the
revised megamaser distance to NGC 4258 of H13 lowers $H_0$ by about $3 \ \kmsMpc$.
Secondly, we have argued that the strong metallicity dependence of the global fits is
caused by sub-luminous outliers in the P-L relation, and may be  unphysical.
Imposing a strong metallicity prior centred
around $Z_W = 0$ raises the LMC solutions for $H_0$ substantially. A
further difference between R11 results and the results presented here 
is that our errors on $H_0$ are larger. This is most noticeable for the MW solutions in Table 3.

 To account
for correlated errors between ground-based and HST photometry and 
correlated errors between the SNe magnitudes, I minimise:
\begin{eqnarray}
\chi^2 &=& \sum_{ij, j=1-9} { (m_{W,ij} - m^P_{W, i})^2 \over (\sigma^2_{e,ij} + \sigma^2_{\rm int})}
+    (m_{W,i} - m^P_{W})(C^{\rm LMC+MW})^{-1}_{ij} (m_{w, j} - m^P_{W}) + (m_{V,i} - m^P_{V})(C^{\rm SNe})^{-1}_{ij} (m_{V, j} - m^P_{V})   \nonumber \\
 & & \qquad +  {(\mu_{0, 4258} - \mu^M_{0, 4258})^2  \over \sigma^2_{\mu_{0, 4248}}}  +
 {(\mu_{0, {\rm LMC}} - \mu^M_{0,{\rm  LMC}})^2  \over \sigma^2_{\mu_{0, {\rm LMC}}}}.
\end{eqnarray}
The first term is summed over the R11 Cepheids as in equation (6). The second term is summed
over the LMC and MW Cepheids, where the covariance matrix $C^{\rm LMC, MW}$ is
\begin{equation}
C^{\rm LMC, MW}_{ij} = (\sigma^2_i + \sigma^2_{\rm int}) \delta_{ij} + \sigma^2_{\rm cal},
\end{equation}
where $\sigma_i$ is the magnitude error, $\sigma_{\rm int}$ is the internal scatter,
and $\sigma_{\rm cal}$ is the calibration error between the ground based and WFC3  
photometry (assumed to be $\sigma_{\rm cal} = 0.04$ mag. as in R11). The third
term is summed over the SNe magnitudes $m_{V, i}$ and the covariance matrix $C^{\rm SNe}$
is
\begin{equation}
C^{\rm SNe}_{ij} = (\sigma^2_i ) \delta_{ij} + \sigma^2_{5a_V},
\end{equation}
where $\sigma_i$ is the SNe magnitude error and $\sigma_{5a_V}$ is the 
error in $5a_V$,  where $a_V$ is  the intercept of the SNe Ia magnitude-redshift relation
($\sigma_{5a_V} = 0.01005$, from R11). Finally, $\mu_{0, 4258}$ and $\mu_{0, \rm LMC}$
are the distance moduli of NGC 4258 and the LMC,  $\mu^M_{0, 4258}$ and $\mu^M_{0, \rm LMC}$
are their measured values (from the maser and eclipsing binary distances respectively)
with errors $\sigma_{\mu_{0, {\rm 4258}}}$ and $\sigma_{\mu_{0, {\rm LMC}}}$. The theoretically
predicted magnitudes are 
\beglet
\begin{eqnarray}
{\rm SNe \ hosts:} & &  m^P_{ij} = \mu_{0, i} + M_W + b_W ({\rm log} P_{ij} -1) + Z_W \Delta {\rm log} {\rm (O/H)}_{ij}, \\
\qquad {\rm NGC \ 4258:} & &  m^P_{j} = \mu_{0, 4258} + M_W + b_W ({\rm log} P_{j} -1) + Z_W \Delta {\rm log} {\rm (O/H)}_j, \\
{\rm LMC:} & &  m^P_{j} = \mu_{0, LMC} + M_W + b_W ({\rm log} P_{j} -1) + Z_W \Delta {\rm log} {\rm (O/H)}_j, \\
{\rm MW:} & &  m^P_{j} =  M_W + b_W ({\rm log} P_{j} -1) + Z_W \Delta {\rm log} {\rm (O/H)}_j,  \\
{\rm SNe:} & &  m^P_{Vi} =  \mu_{0, i} + 5 {\rm log} H_0 - 25 - 5a_V.
\end{eqnarray}
\endlet

Results for joint distance anchor fits are given in Table 4. There are several points worth 
noting:

\smallskip

\noindent
$\bullet$ As in the analyses presented in  previous Sections, the low metallicity Cepheid
outliers in the R11 data lead to a sensitivity to the outlier rejection criterion.  

\smallskip

\noindent
$\bullet$ With no metallicity or slope priors, the combined solutions using NGC 4258 + LMC anchors
are discrepant at $\simgt 2\sigma$ with the MW anchor solutions given in Table 3. (Compare,
for example, the results for fits 33 and 45.)

\smallskip

\noindent
$\bullet$ Applying a strong metallicity prior, the solutions for $H_0$ for the combined
NGC 4258 and LMC fits  become insensitive
to the outlier rejection criteria (and are consistent to within $\sim 0.5 \ \kmsMpc$).

\smallskip

\noindent
$\bullet$ Since the slope of the P-L relation is well constrained by the LMC Cepheids,
adding a slope prior has almost no effect on  solutions that include the LMC Cepheids.

If we accept each of these distance anchors at face value, and average over the 
outlier rejection criteria as in the previous Section using the solutions
with strong metallicity and no slope priors for (18a)  and (18d)
and strong metallicity and slope priors for (18b) and  (18c), then we find:
\beglet
\begin{eqnarray}
H_0 &=& 71.8  \pm  2.6 \ \kmsMpc,   \quad {\rm NGC \ 4258+LMC},  \label{H04258LMC} \\
H_0 &=& 72.2  \pm  2.8 \ \kmsMpc,   \quad {\rm NGC \ 4258+MW},  \label{H04258MW} \\
H_0 &=& 73.9  \pm  2.7 \ \kmsMpc,   \quad {\rm LMC+MW},  \label{H0LMCMW} \\
H_0 &=& 72.5  \pm  2.5 \ \kmsMpc,   \quad {\rm NGC \ 4258+LMC+MW}.  \label{H04258LMCMW}
\end{eqnarray}
\endlet
These values differ  by $1.6\sigma$, $1.6\sigma$, $2.2\sigma$ and  $1.9 \sigma$ respectively from the \Planck\ value of $H_0$ for the base \lcdm\ model. Evidently, using the LMC and especially the 
MW Cepheids as distance
anchors pulls $H_0$ to higher values than those derived using the megamaser distance. 
Note that when R11 combine all three distance anchors, they conservatively adopt the largest error
from any pair of distance anchors. Adopting the same approach would increase the error in (18d) to
$2.8 \ \kmsMpc$.

\subsection{Consistency of distance anchors}

Before accepting (18a) - (18d) it is worth investigating internal
consistency tests of the three distance anchors. We first use the MW
Cepheids to compute the distance modulus to the LMC assuming no
metallicity dependence of the P-L relation.  The result is
\begin{equation}
\mu_{0, LMC} = 18.455 \pm 0.042,
\end{equation} 
in good agreement with the Pietrzy\'nski \etals
(2013) eclipsing binary distance modulus of $18.493 \pm 0.049$. In fact,
combining these estimates we deduce
\begin{equation}
Z_W = 0.10 \pm 0.16 \ {\rm mag.} \ {\rm dex}^{-1},
\end{equation}
consistent with a weak metallicity dependence of the P-L relation at H-band.

\begin{figure}

\begin{center}
\includegraphics[width=120mm, angle=0]{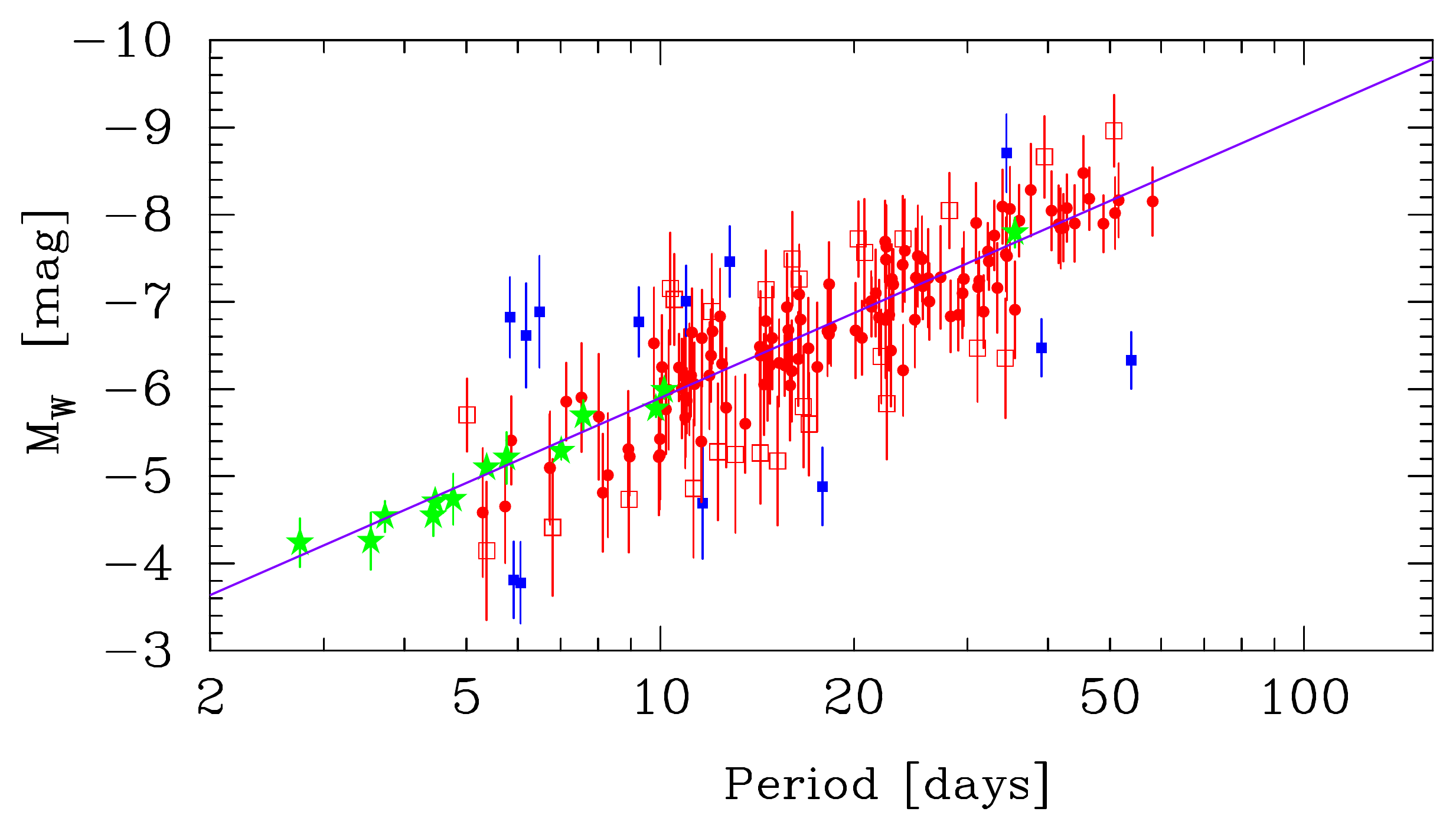}
\end{center}
\caption
{Period-luminosity fit used to determine the distance modulus to NGC 4258
using the MW Cepheids as a distance anchor. The MW Cepheids are
shown by the (green) filled stars. The rest of the points (red and blue) show the R11  Cepheids.
As in Figure 2, filled (red) circles are accepted by both the
R11 and $T=2.25$ rejection criteria while filled (blue) sqaures are rejected
by both criteria. Open (red) squares are rejected by R11 but accepted by the $T=2.25$ criterion. The line shows the best fit P-L relation.}. 
\end{figure}

Next, we use the LMC Cepheids to determine a distance modulus to NGC
4258, assuming the Pietrzy\'nski \etals (2013) eclipsing binary
distance. This solution is based on the likelihood of equation
(14). I impose the strong metallicity prior and average over the three
outlier rejection criteria, adding the scatter to the final error
estimate. The result is
\begin{equation}
\mu_{0, 4258} = 29.285 \pm 0.083,  \label{distmod1}
\end{equation} 
which is within $1.1\sigma$ of the H13 megamaser distance modulus of 
$29.404\pm 0.066$. Combining these estimates we deduce
\begin{equation}
Z_W = -0.29 \pm 0.26 \ {\rm mag.} \ {\rm dex}^{-1},
\end{equation}
consistent with zero but with even lower precision than the estimate of
(20).

Finally, I use the MW Cepheids to compute a distance modulus to NGC
4258. Since the MW Cepheids have similar metallicities to the mean of
the Cepheids in NGC 4258, uncertainties in the metallicity dependence
of the P-L relation do not introduce a significant source of error
into the distance modulus. Nevertheless, I impose the strong
metallicity prior on the solutions and average over the three outlier
criteria. This gives
\begin{equation}
\mu_{0, 4258} = 29.241 \pm 0.079,   \label{distmod2}
\end{equation} 
which is about $1.6 \sigma$ lower than the H13 distance modulus. Figure 5
shows  the $T=2.25$ fit to the MW and NGC 4258 Cepheids.

The differences between the $H_0$ values for the three distance
anchors are reflected by these differences in the distance moduli of
NGC 4258. The MW and LMC Cepheids (assuming zero metallicity
dependence) give a shorter distance to NGC 4258 than the H13 revised megamaser
distance. It is possible that the true distance to NGC 4258 is
substantially lower than the H13 central value. However, it is also possible that
the tension is caused by a more subtle effect, for example, a residual
$\sim 0.1$ mag.  bias of the NGC 4258 P-L relation caused by
asymmetric outliers or systematic errors in the corrections for
crowding biases.

Further work on the tension between (\ref{distmod1}), (\ref{distmod2})
and the H13 distance modulus is required to improve on the estimate of
(\ref{H04258LMCMW}). Evidently, the discrepancies are not of high enough
statistical significance to reveal an obvious inconsistency between
any of the distance anchors. However, the difference between (23) and
the H13 distance modulus is quite large and suggestive of a possible
problem.

\section{Conclusions and Discussion}

\begin{figure}

\begin{center}
\includegraphics[width=90mm, angle=0]{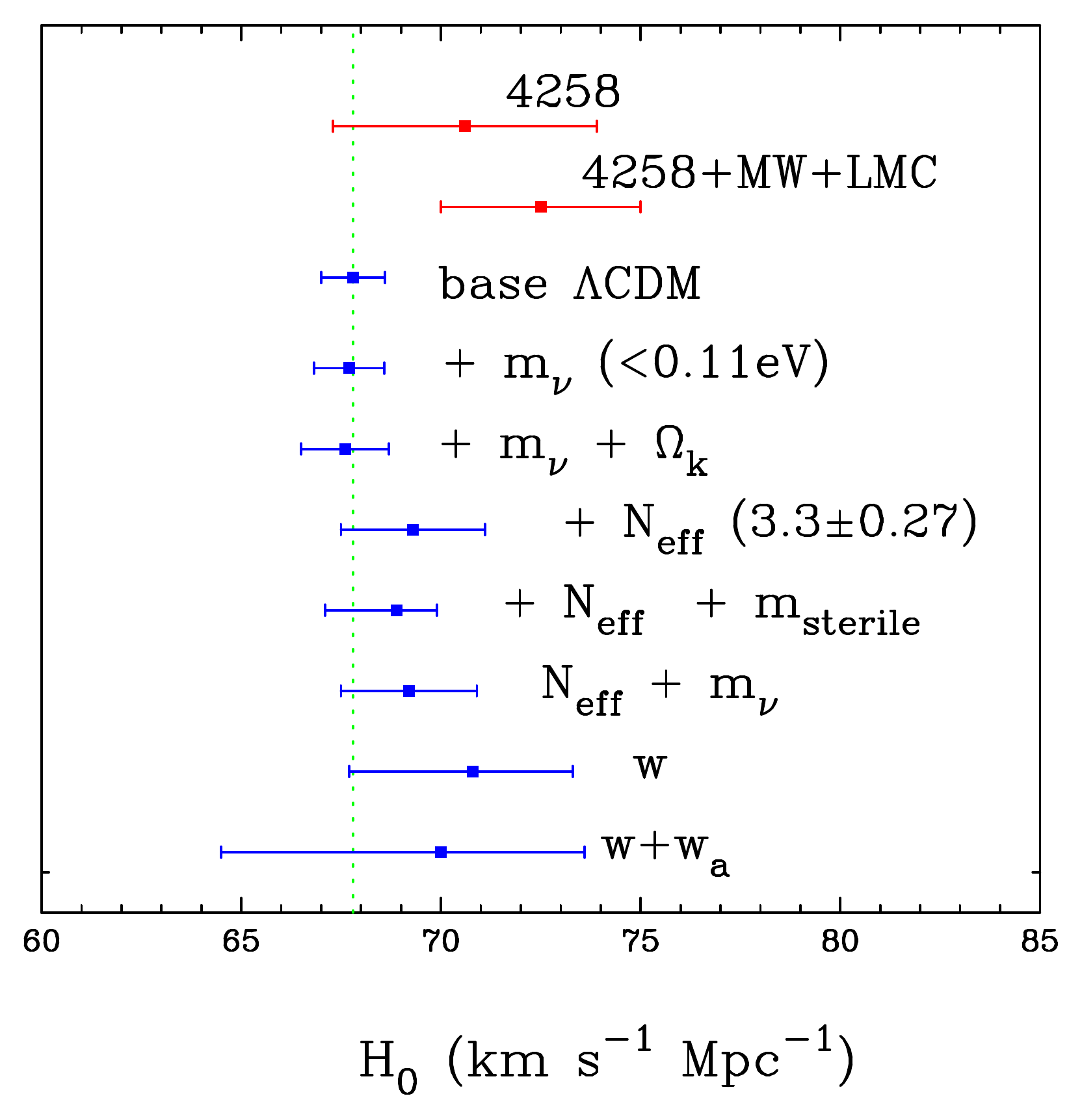}
\end{center}
\caption
{The direct estimates (red) of $H_0$  (together with $1\sigma$
error bars) for the NGC 4258 distance anchor (equation 9) and for all three distance anchors (equation 18d). The remaining (blue)  points show the constraints from P13
for the base \LCDM\ cosmology and some extended models combining CMB data
with data from baryon acoustic oscillation surveys. The extensions are
as follows: $m_\nu$, the mass of a single neutrino species; $m_\nu + \Omega_k$,
allowing a massive neutrino species and spatial curvature; $N_{\rm eff}$, allowing
additional relativistic neutrino-like particles; $N_{\rm eff} + m_{\rm sterile}$,
adding a massive sterile neutrino and additional relativistic particles;
$N_{\rm eff} + m_\nu$, allowing a massive neutrino and additional relativistic
particles; $w$, dark energy with a constant equation of state $w = p/\rho$;
$w+w_a$ , dark energy with a time varying equation of state. I give the
$1\sigma$ upper limit on $m_\nu$ and the $1\sigma$ range for $N_{\rm eff}$. See P13 for
further details on these extended models.}
\end{figure}

The SH0ES project was cleverly designed to minimise the impact of
metallicity, crowding, and photometric calibration biases when
comparing Cepheids measured in SNe hosts with those in NGC 4258.
These methodological reasons argue that higher weight should be placed
on the NGC 4258 anchor than either the LMC or MW anchors when using
the R11 data. However, the value of $H_0$ derived using NGC 4258 as an
anchor relies on the fidelity of the geometric maser distance. Despite
the extensive VLBI campaign described by H13, systematic errors
contribute significantly to the total error in the megamaser
distance. It may therefore be dangerous to place very high weight on
the NGC 4258 distance without cross-checks with independent distance
anchors.

However, in addition to the methodological issues that drove the
design of the SH0ES project, there are significant problems in using
the LMC and MW distance anchors.  Although there are several accurate
and consistent eclipsing binary distance estimates to the LMC
(Fitzpatrick \etals 2002; Ribas \etals 2002; Pietrzy\'nski \etals
2009, 2013) $H_0$ derived using the LMC as a distance anchor is
extremely sensitive to any metallicity dependence of the P-L relation
(equation 11). I show that the R11 sample contains sub-luminous low
metallicity Cepheids, pointing either to a stronger than expected
metallicity dependence of the near-IR P-L relation (in conflict with
the Freedman and Madore 2011 analysis), a possible misidentification
of these objects as classical Cepheids, or to some unidentified
systematic error in their magnitudes. The presence of these
sub-luminous Cepheids causes some sensitivity to the rejection
criteria used to identify outliers from the mean P-L
relation. However, if a strong metallicity prior is imposed, the
global fits and derived values of $H_0$ become insensitive to the
outlier rejection criteria. (It is also worth noting that the strong
metallicity prior also affects the value of $H_0$ derived using NGC
4258 as a distance anchor: we find $H_0 = 70.6 \pm 3.3 \ \kmsMpc$
compared to the value $H_0 = 72.0 \pm 3.0 \ \kmsMpc$ quoted by H13.)
One would have greater confidence in using the LMC anchor if there
were stronger observational constraints on $Z_W$.

The sample of MW Cepheids with parallax measurements is small and
contains only one star that overlaps with the period range sampled by
Cepheids in the SNe host galaxies ({\it cf} Figures 2 and 4). Use of
the MW Cepheids as an anchor is therefore susceptible to sample biases
and small number statistics.  The distance modulus to NGC 4258 derived
from the MW Cepheids is lower by about $1.6 \sigma$ compared to the
H13 megamaser distance modulus.  As a consequence, $H_0$ derived using
the MW Cepheids as a distance anchor is higher than that derived from
the megamaser distance and is discrepant by about $2.2\sigma$ with the
\planck\ base \LCDM\ value. This is the largest discrepancy
reported in this paper with the Planck determination of $H_0$.
Observations with the GAIA satellite will increase
the number of Galactic Cepheids with accurate parallaxes into the many
thousands. It will be interesting to see whether the tensions with the
megamaser distance and with the \planck\ base \LCDM\ cosmology persist.

The value of $H_0$ derived here from the megamaser distance is within
$1\sigma$ of the \planck\ base \LCDM\ value of $H_0$. Although there are
some tensions between the three distance anchors, none are sufficiently
compelling to justify excluding either the MW or LMC anchors from a
joint fit. Imposing the strong metallicity prior, the combination of
all three distance anchors raises $H_0$ to $72.5 \pm 2.5 \ \kmsMpc$,
which is within $1.9 \sigma$ of the \planck\ base \LCDM\ value.

Figure 6 compares these two estimates of $H_0$ with the P13 results
from the \planck+WP+highL+BAO\footnote{\planck\ temperature likelihood
  combined with the WMAP polarization likelihood at low multipoles
  combined with high resolution CMB experiments combined with baryon
  acoustic oscillation (BAO) measurements.}  likelihood for the base
\LCDM\ cosmology and some extended \LCDM\ models.  I show the
combination of CMB and BAO data since $H_0$ is poorly constrained for
some of these extended models using CMB temperature data alone.  (For
reference, for this data combination $H_0 =67.80 \pm 0.77 \
\kmsMpc$ in the base \LCDM\ model.) 
The combination of CMB and BAO data is certainly not prejudiced
against new physics, yet the $H_0$ values for the extended \LCDM\
models shown in this figure all lie within $1\sigma$ of the best fit
value for the base \LCDM\ model. For example, in the models exploring
new physics in the neutrino sector, the central value of $H_0$ never
exceeds $69.3 \ \kmsMpc$. If the true value of $H_0$ lies closer to,
say, $H_0 = 74 \ \kmsMpc$, the dark energy sector, which is poorly
constrained by the combination of CMB and BAO data, seems a more
promising place to search for new physics.  

In summary, the discrepancies between the \planck\ results and the
direct $H_0$ measurements shown in Figure 5  are not large enough to
provide compelling evidence for new physics beyond the base \LCDM\
cosmology.

\medskip

\noindent {\bf Acknowledgements:} I am particularly grateful to Adam
Riess, who has answered patiently my many questions on the analysis of
the SH0ES data. I also thank Rob Kennicutt, Wendy Freedman and Brian
Schmidt for valuable discussions and correspondence and the referee for
a helpful report.

\appendix
\section{Extending the period range of the global fits}

Throughout this paper, I imposed an upper period limit of 60 days on
the R11 Cepheid sample.  As noted in Section 2, there is evidence from
analyses of LMC Cepheids that the P-L relation flattens for Cepheids
with periods $>60$ days (Persson \etal, 2004; Freedman \etals 2011;
Scowcroft \etal 2011). R11, however, used all Cepheids with periods $<
205$ days in their analysis. The purpose of this Appendix is to show
how the results of Sections 3 and 4 change if the Cepheid period range
is extended.  Table A1 is the equivalent of Tables 2 and 3, but using
R11 Cepheids with periods $<205$ days.  The main change is that the
slopes of the P-L relation in many of the fits become substantially
flatter than the LMC slopes of equation (4). As a consequence, the
global fits in Table A1 become more sensitive to the imposition of an
LMC slope prior, whereas the fits in Tables 2 and 3 are insensitive to
the slope prior. This is particularly true for fits A25-A36 using the
MW Cepheids as an anchor. Without any slope prior, $H_0$ is about $80
\ \kmsMpc$ {\it and is inconsistent with $H_0$ determined using the maser
distance to NGC 4258} (fits A1-A12). This provides further evidence
that  the  MW Cepheids are the least reliable of the three distance anchors.
R11 adopted a slope prior of $-3.3 \pm 0.1$ in their fits involving MW Cepheids. This
prior is, however, inconsistent with the shallower slopes of the SNe host galaxy
P-L relations derived using Cepheids with periods up to $205$ days.

Comparing Table A1 with Tables 2 and 3, we see that the flatter slopes of the P-L relation 
have little impact on $H_0$ using either NGC 4258 or the LMC Cepheids as a distance anchor.
However, adopting an  upper period of limit of 60 days,  as in the main body of this paper, 
all of the P-L slopes are consistent with the LMC P-L relation. The global fits
are then insensitive to the imposiition of an LMC slope prior (even for the MW Cepheids).

\begin{table*}

   \center{Global fits: NGC 4258 anchor \hfill}
\begin{center}
{\scriptsize \begin{tabular}{llllllllll} \hline
               &            \multicolumn{7}{c}{T=2.5 Rejection} & \multicolumn{2}{c}{Priors} \\
Fit        & $N_{\rm fit}$ & $\hat \chi^2_{\rm WFC3}$ & $H_0$ & $p_W$ & $b_W$ & $Z_w$ & $\sigma_{\rm int}$& $Z_w$& $b_W$   \\ \hline
A1    & $552$ & $1.00$ & 71.5 (3.1) (2.2)  & 26.08 (0.12)  & -2.88 (0.08)  & -0.33 (0.14)  & 0.30  &  N & N  \\
A2    & $550$ & $1.00$ & 71.6 (3.1) (2.2)  & 26.09 (0.12)  & -2.88 (0.08)  & -0.21 (0.12)  & 0.29  &  W & N  \\
A3    & $551$ & $1.00$ & 71.0 (3.0) (2.1)  & 26.12 (0.12)  & -2.91 (0.08)  & -0.005 (0.020)  & 0.30  &  S & N  \\
A4    & $551$ & $1.00$ & 69.2 (2.9) (2.0)  & 26.33 (0.09)  & -3.06 (0.06)  & -0.004 (0.020)  & 0.31  &  S & Y  \\ \hline

               &            \multicolumn{7}{c}{T=2.25 Rejection} & \multicolumn{2}{c}{Priors} \\
Fit        & $N_{\rm fit}$ & $\hat \chi^2_{\rm WFC3}$ & $H_0$ & $p_W$ & $b_W$ & $Z_w$ & $\sigma_{\rm int}$& $Z_w$& $b_W$   \\ \hline
A5    & $523$ & $1.00$ & 72.0 (2.9) (2.0)  & 26.21 (0.10)  & -2.95 (0.07)  & -0.46 (0.12)  & 0.21  &  N & N  \\
A6    & $524$ & $1.00$ & 71.7 (2.9) (1.9)  & 26.22 (0.10)  & -2.96 (0.07)  & -0.32 (0.11)  & 0.21  &  W & N  \\
A7    & $524$ & $1.00$ & 71.3 (2.9) (1.9)  & 26.20 (0.10)  & -2.95 (0.07)  & -0.006 (0.020)  & 0.21  &  S & N  \\
A8    & $519$ & $1.00$ & 70.9 (2.9) (1.9)  & 26.38 (0.09)  & -3.06 (0.06)  & -0.006 (0.020)  & 0.20  &  S & Y  \\ \hline

               &            \multicolumn{7}{c}{R11 Rejection} & \multicolumn{2}{c}{Priors} \\
Fit        & $N_{\rm fit}$ & $\hat \chi^2_{\rm WFC3}$ & $H_0$ & $p_W$ & $b_W$ & $Z_w$ & $\sigma_{\rm int}$& $Z_w$& $b_W$   \\ \hline
A9    & $448$ & $0.65$ & 72.3 (2.8) (1.8)  & 26.30 (0.10)  & -3.01 (0.07)  & -0.26 (0.10)  & $0.21$  &  N & N  \\
A10    & $448$ & $0.65$ & 72.1 (2.8) (1.7)  & 26.31 (0.10)  & -3.01 (0.07)  & -0.21 (0.07)  & $0.21$  &  W & N  \\
A11    & $448$ & $0.66$ & 71.4 (2.7) (1.7)  & 26.35 (0.07)  & -3.04 (0.07)  & -0.006 (0.016)  & $0.21$  &  S & N  \\
A12    & $448$ & $0.66$ & 70.7 (2.7) (1.6)  & 26.45 (0.07)  & -3.11 (0.05)  & -0.006 (0.016)  & $0.21$  &  S & Y  \\ \hline

\end{tabular}}

\medskip

           Global fits: LMC anchor

{\scriptsize\begin{tabular}{llllllllll} \hline
               &            \multicolumn{7}{c}{T=2.5 Rejection} & \multicolumn{2}{c}{Priors} \\
Fit        & $N_{\rm fit}$ & $\hat \chi^2_{\rm WFC3}$ & $H_0$ & $p_W$ & $b_W$ & $Z_w$ & $\sigma_{\rm int}$& $Z_w$& $b_W$   \\ \hline
A13    & $553$ & $1.00$ & 71.3 (3.6) (2.2)  & 15.54 (0.08)  & -3.11 (0.04)  & -0.27 (0.14)  & 0.31  &  N & N  \\
A14    & $552$ & $1.00$ & 72.1 (3.4) (2.1)  & 15.57 (0.07)  & -3.11 (0.04)  & -0.20 (0.12)  & 0.31  &  W & N  \\
A15    & $551$ & $1.00$ & 74.2 (2.9) (1.9)  & 15.65 (0.05)  & -3.11 (0.04)  & -0.003 (0.020)  & 0.31  &  S & N  \\
A16    & $551$ & $1.00$ & 73.9 (2.9) (1.8)  & 15.67 (0.05)  & -3.13 (0.04)  & -0.003 (0.020)  & 0.31  &  S & Y  \\ \hline

               &            \multicolumn{7}{c}{T=2.25 Rejection} & \multicolumn{2}{c}{Priors} \\
Fit        & $N_{\rm fit}$ & $\hat \chi^2_{\rm WFC3}$ & $H_0$ & $p_W$ & $b_W$ & $Z_w$ & $\sigma_{\rm int}$& $Z_w$& $b_W$   \\ \hline
A17    & $523$ & $1.00$ & 69.9 (3.2) (2.0)  & 15.49 (0.07)  & -3.11 (0.04)  & -0.38 (0.12)  & 0.21  &  N & N  \\
A18    & $523$ & $1.00$ & 70.5 (3.2) (2.0)  & 15.52 (0.07)  & -3.11 (0.04)  & -0.31 (0.11)  & 0.21  &  W & N  \\
A19    & $519$ & $1.00$ & 74.2 (2.8) (1.8)  & 15.65 (0.05)  & -3.11 (0.04)  & -0.006 (0.020)  & 0.20  &  S & N  \\
A20    & $518$ & $1.00$ & 73.9 (2.8) (1.7)  & 15.68 (0.05)  & -3.14 (0.04)  & -0.006 (0.020)  & 0.20  &  S & Y  \\ \hline

               &            \multicolumn{7}{c}{R11 Rejection} & \multicolumn{2}{c}{Priors} \\
Fit        & $N_{\rm fit}$ & $\hat \chi^2_{\rm WFC3}$ & $H_0$ & $p_W$ & $b_W$ & $Z_w$ & $\sigma_{\rm int}$& $Z_w$& $b_W$   \\ \hline
A21    & $448$ & $0.65$ & 72.1 (3.1) (1.7)  & 15.58 (0.06)  & -3.13 (0.03)  & -0.22  (0.10)  & $0.21$  &  N & N  \\
A22    & $448$ & $0.65$ & 72.5 (3.0) (1.7)  & 15.60 (0.06)  & -3.13 (0.03)  & -0.18 (0.09)  & $0.21$  &  W & N  \\
A23    & $448$ & $0.66$ & 74.2 (2.7) (1.7)  & 15.68 (0.04)  & -3.14 (0.03)  & -0.005 (0.016)  & $0.21$  &  S & N  \\
A24    & $448$ & $0.66$ & 73.9 (2.6) (1.5)  & 15.70 (0.04)  & -3.15 (0.03)  & -0.005 (0.016)  & $0.21$  &  S & Y  \\ \hline

\end{tabular}}

\medskip

             Global fits: MW Cepheids anchor
{\scriptsize \begin{tabular}{llllllllll} \hline
               &            \multicolumn{7}{c}{T=2.5 Rejection} & \multicolumn{2}{c}{Priors} \\
Fit        & $N_{\rm fit}$ & $\hat \chi^2_{\rm WFC3}$ & $H_0$ & $M_W$ & $b_W$ & $Z_w$ & $\sigma_{\rm int}$& $Z_w$& $b_W$   \\ \hline
A25    & $552$ & $1.00$ & 82.0 (3.9) (3.6)  & -5.85 (0.05)  & -2.96 (0.08)  & -0.31 (0.13)  & 0.31  &  N & N  \\
A26    & $549$ & $1.00$ & 82.1 (3.9) (3.5)  & -5.85 (0.05)  & -2.96 (0.08)  & -0.20 (0.12)  & 0.29  &  W & N  \\
A27    & $551$ & $1.00$ & 80.2 (3.7) (3.4)  & -5.85 (0.05)  & -3.00 (0.08)  & -0.004 (0.020)  & 0.31  &  S & N  \\
A28    & $551$ & $1.00$ & 77.7 (3.4) (3.0)  & -5.87 (0.05)  & -3.08 (0.06)  & -0.006 (0.020)  & 0.30  &  S & Y  \\ \hline

               &            \multicolumn{7}{c}{T=2.25 Rejection} & \multicolumn{2}{c}{Priors} \\
Fit        & $N_{\rm fit}$ & $\hat \chi^2_{\rm WFC3}$ & $H_0$ & $M_W$ & $b_W$ & $Z_w$ & $\sigma_{\rm int}$& $Z_w$& $b_W$   \\ \hline
A29    & $522$ & $1.00$ & 81.2 (3.6) (3.3)  & -5.86 (0.05)  & -2.99 (0.06)  & -0.45 (0.12)  & 0.20  &  N & N  \\
A30    & $521$ & $1.00$ & 80.9 (3.6) (3.3)  & -5.86 (0.05)  & -2.99 (0.07)  & -0.36 (0.11)  & 0.20  &  W & N  \\
A31    & $514$ & $1.00$ & 81.2 (3.5) (3.2)  & -5.85 (0.06)  & -2.96 (0.06)  & -0.007 (0.020)  & 0.18  &  S & N  \\
A32    & $519$ & $1.00$ & 77.8 (3.3) (2.9)  & -5.87 (0.05)  & -3.08 (0.05)  & -0.006 (0.020)  & 0.20  &  S & Y  \\ \hline

               &            \multicolumn{7}{c}{R11 Rejection} & \multicolumn{2}{c}{Priors} \\
Fit        & $N_{\rm fit}$ & $\hat \chi^2_{\rm WFC3}$ & $H_0$ & $M_W$ & $b_W$ & $Z_w$ & $\sigma_{\rm int}$& $Z_w$& $b_W$   \\ \hline
A33    & $448$ & $0.66$ & 79.9 (3.2) (2.8)  & -5.87 (0.04)  & -3.05 (0.06)  & -0.25 (0.11)  & $0.21$  &  N & N  \\
A34    & $448$ & $0.66$ & 79.5 (3.2) (2.8)  & -5.87 (0.04)  & -3.06 (0.06)  & -0.20 (0.10)  & $0.21$  &  W & N  \\
A35    & $448$ & $0.67$ & 78.2 (3.1) (2.7)  & -5.87 (0.04)  & -3.08 (0.06)  & -0.006 (0.016)  & $0.21$  &  S & N  \\
A36    & $448$ & $0.67$ & 76.6 (2.8) (2.4)  & -5.87 (0.04)  & -3.14 (0.05)  & -0.005 (0.016)  & $0.21$  &  S & Y  \\ \hline

\end{tabular}}
\end{center}

\smallskip

\caption{Fits as in  Tables 2 and 3, but using all Cepheids in the R11 sample with periods
less than 205 days.  The columns are as defined in Table 2.}

\end{table*}

\end{document}